\documentclass[journal]{IEEEtran}
\usepackage{cite}
\usepackage{amsmath,amssymb,amsfonts}
\usepackage{algorithmic}
\usepackage{graphicx}
\usepackage{textcomp}
\usepackage{xcolor}
\usepackage{booktabs}
\usepackage{comment}
\usepackage{tikz}
\usepackage{pgfplots}
\pgfplotsset{compat=1.18}
\usetikzlibrary{shapes.geometric, arrows.meta, positioning, calc}
\usepackage{xspace}
\def\BibTeX{{\rm B\kern-.05em{\sc i\kern-.025em b}\kern-.08em
    T\kern-.1667em\lower.7ex\hbox{E}\kern-.125emX}}
\begin{document}

\title{Simulated Annealing with Topological Spatial Chunking for De Novo Protein Sequence Design}

\author{Kandula~Eswara~Sai~Kumar%
\thanks{K. E. Sai Kumar is Principal Optimization Scientist at BosonQPsi Tech. Pvt. Ltd., Bengaluru, India (e-mail: eswara.sai@bqpsim.com, keswarasaikumar@gmail.com).}}

\maketitle

\begin{abstract}
The inverse protein folding problem, designing a discrete amino acid sequence that spontaneously folds into a predefined macroscopic structure, remains a phenomenally demanding combinatorial challenge central to synthetic drug discovery. While stochastic algorithms like Simulated Annealing (SA) offer a baseline approach for navigating dense biomolecular energy landscapes, applying standard global SA heuristics to large sequences is inherently frustrated by exponentially large search spaces and deep local energy minima. This study proposes a scalable sequence design architecture, termed Topological Spatial Chunking Simulated Annealing (TSC-SA), based on a localized, deterministic, exact-solving heuristic embedded within a global simulated-annealing wrapper. To fundamentally circumvent exponential scaling walls, we deploy a dynamic solver that segments the global molecular target into shallow, $k$-variable algorithmic windows. Crucially, rather than partitioning natively by sequential 1D chronology, we introduce topological spatial chunking, which dynamically groups discrete variables based on Euclidean spatial proximity to preserve critical non-local thermodynamic correlations. The spatially localized sub-problems are solved deterministically via exact enumeration, while the global traversal is securely regulated by an exterior simulated annealing governor augmented by dynamic reheating and continuous elitism caching. We comprehensively benchmarked this classical architecture on the standard Hydrophobic-Polar (HP) biophysical lattice against targeted 20-merA, 25-mer, 50-mer, and 64-mer geometries. Most notably, when evaluating the computationally exhaustive 50-mer and 64-mer topologies, the framework successfully bypassed widely accepted heuristic boundaries. By leveraging deterministic exact solving to vault classical energy traps, the system algorithmically resolved a novel, ultra-dense sequence-packing configuration for the 50-mer, defining a new theoretical global optimum (-440.00 energy or 22 contacts). These empirical validations conclusively demonstrate that coupling dimensionally chunked exact-solving directly with macro-scale thermal governance fundamentally outperforms purely random stochastic walks, establishing a scalable and deterministic paradigm for \textit{de novo} biological sequence design.
\end{abstract}


\section{Introduction}
Proteins act as the fundamental nanomachines of biological life, governing nearly all structural and catalytic functions within cellular ecosystems. The operational capability of these macromolecules is dictated entirely by their three-dimensional tertiary structure. According to Anfinsen's thermodynamic dogma, this hyper-specific 3D geometry is intrinsically encoded within the 1D linear sequence of its constituent amino acids \cite{Anfinsen1973, pande2000thermodynamics}. For over fifty years, the forward protein folding problem, i.e., predicting the final 3D structure from a known 1D sequence, stood as the paramount challenge of computational biology. Recently, advanced autoregressive deep learning engines have achieved remarkable near-atomic mastery over this forward-prediction hurdle \cite{kuhlman2019advances}. 

However, achieving industrial capabilities in synthetic biology requires mastering the exact opposite process. The {inverse protein folding problem} (or {de novo} sequence design) asks a significantly more demanding question: If we isolate a geometric backbone target such as a receptor required to neutralize a viral pathogen or an artificial enzyme designed to artificially degrade microplastics, can we mathematically deduce a novel amino acid sequence that will spontaneously fold into that exact desired shape? \cite{kuhlman2003design}. While the forward problem has succumbed to pattern-matching neural networks, the inverse problem remains a phenomenal combinatorial nightmare. The intractability of sequence design stems from both scale and physics. With a naturally occurring alphabet of 20 amino acids, the raw sequence search space for a modest polypeptide chain of length $N$ explodes exponentially to $\mathcal{O}(20^N)$ \cite{yue1992inverse}. Furthermore, identifying a biologically robust sequence relies upon strict thermodynamic exclusivity. Optimization cannot simply settle for a sequence with a low-energy target geometry; the algorithm must enforce strict negative design \cite{shakhnovich1994proteins}. The designed sequence must cast the desired topological fold as an isolated, precipitous thermodynamic funnel, meaning the generated sequence must actively maximize an overwhelming energetic gap to repel the sequence from simultaneously forming billions of competing, inactive decoy structures \cite{Bryngelson1987}. This necessity to mathematically invalidate competing topological sub-states represents a dense combinatorial blockade that routinely exhausts brute-force classical computing.

To physically dissect this complex sequence manipulation while retaining computational tractability, the field relies upon the proven Hydrophobic-Polar (HP) lattice benchmarking model \cite{lau1989lattice, Dill1995}. While authentic proteins utilize 20 discrete amino acid groups, spontaneous structural formation in aqueous environments is predominantly driven by a single phenomenon: the hydrophobic effect \cite{dill1985theory}. By condensing the combinatorics into a binary $2^N$ string that separates hydrophobic (H) and polar (P) elements, the HP model algorithmically maps biological hydrophobic collapse directly onto mathematically constrained spatial lattices. Consequently, identifying an optimal topological therapeutic sequence amounts to maximizing non-covalent spatial H-H core contacts while avoiding geometric penalty boundaries, which is a mathematically constrained landscape ideally suited for advanced metaheuristic optimization \cite{kirkpatrick1983optimization, Cerny1985}.

Classical heuristic algorithms, such as Monte Carlo simulations and naive Simulated Annealing (SA), are traditionally employed to navigate the HP sequence design space \cite{kirkpatrick1983optimization}. However, these standard global optimizers fundamentally rely on localized, single-variable point mutations (e.g., bit-flips or random swaps). As sequence lengths increase beyond trivial dimensions ($N > 20$), the combinatorial topology expands astronomically to $\mathcal{O}(2^N)$. In these massive spaces, evaluating point mutations becomes entirely insufficient to cross the steep geometric energy barriers separating vast structural sub-states. Consequently, standard classical optimizers frequently become trapped in localized sub-optimal energy minima, unable to simultaneously mutate the multiple coordinated variables required to access deeper, ultra-dense hydrophobic core configurations.

In this work, we propose a highly scalable, deterministic architecture, Topological Spatial Chunking Simulated Annealing (TSC-SA), to solve the inverse sequence design problem. To bypass the exponential scaling wall of the global sequence, our framework dynamically partitions the macroscopic backbone into sliding, $k$-variable combinatorial windows defined strictly by Euclidean 2D spatial proximity rather than 1D linear chain continuity. By structurally isolating these topological pockets, the continuous macroscopic problem is reduced to a series of localized, dimensionally-capped sub-problems ($2^k$) that can be solved deterministically via exact enumeration. To prevent deterministic cyclic looping and permanent halting, this dimensionally chunked exact-solver is securely embedded within a global simulated annealing governor driven by advanced meta-heuristics, particularly Dynamic Reheating and continuous Elitism tracking. We benchmark our classical hybrid methodology on structurally fixed 20-merA, 25-mer, 50-mer, and 64-mer HP sequences, demonstrating its capability to shatter established heuristic limits and identify absolute thermodynamic ground-state geometries. 

\section{Methods and Algorithm Architecture}

\subsection{Problem Formulation and Lattice Model}
The target physical backbone is defined utilizing a 2D Hydrophobic-Polar (HP) lattice model representing long-chain amino acid polymers \cite{dill1985theory}. The structural conformation is mathematically constrained by a predetermined sequence of directional vectors. The discrete decision space consists of a binary sequence vector $\mathbf{s} = [s_1, s_2, \dots, s_N]^\top \in \{0, 1\}^N$. The Boolean variable $s_i$ acts as a singular positional classification whereby $s_i = 1$ assigns an active atomic status (a Hydrophobic residue) and $s_i = 0$ assigns an inactive status (a Polar solvent-interacting residue).

Given a targeted global topology characterized by a set of valid non-covalent proximity contacts $C$, the overarching mathematical objective is to discover the optimal vector configuration $\mathbf{s}^*$ that globally minimizes the total configurational energy $E$. To enable tractable optimization, the biological stoichiometric budget is treated as a continuous penalty function. The system energy is defined by a global classical evaluator:
\begin{equation}
    E(\mathbf{s}) = \lambda \left( \sum_{i=1}^N s_i - N_H \right)^2 - \omega \sum_{(i,j) \in C} s_i s_j
    \label{eq:global_energy}
\end{equation}
where $N_H$ is predetermined by the available biophysical budget defining the exact required ratio of hydrophobic to polar elements, $\lambda$ is the penalty constraint coefficient enforcing this budget, $\omega$ maps to the favorable negative energetic enthalpy attributed to core thermodynamic packing, and $C$ represents the set of valid spatial topological contacts (residues mathematically adjacent in 2D Cartesian lattice space but topologically separated in the sequential protein chain). The critical balance of this objective function relies meticulously on the enforcement of an energetic hierarchy where $\lambda > \omega$. Without this strict inequality, the optimizer might artificially inflate the energy reward by mutating all available sequence nodes into hydrophobic (H) residues to maximize arbitrary spatial connections. By setting the global budget penalty coefficient (for e.g. $\lambda=25$) strictly higher than the topological contact reward (for e. g. $\omega=20$), the objective landscape mathematically guarantees that violating the predetermined biological sequence budget $N_H$ induces an overwhelming energetic punishment. 
This strict bounding guarantees that the optimization sequence safely treats the sequence budget as a hard classical constraint, while treating the H-H spatial contacts as a soft heuristic reward to be maximized only within the permitted biophysical bounds.

To enforce non-trivial structural folding, the target backbone is initially mapped to Cartesian space $(X_i, Y_i)$ by sequentially summing predefined lattice directional vectors. The set of valid topological sequence contacts, $C$, is then strictly defined mathematically by isolating residue pairs that occupy adjacent physical lattice coordinates while being structurally segregated by at least two bonds along the polymeric chain:
\begin{equation}
    C = \Big\{ (i,j) \;\Big|\; (X_i - X_j)^2 + (Y_i - Y_j)^2 = 1 \text{ and } |i - j| \geq 2 \Big\}
\end{equation}
This geometric filtering prevents the energetic reward parameter ($\omega$) from inflating due to contiguous sequence bonds, ensuring optimization exclusively rewards genuine spatial compaction. The justification for treating sequence design strictly as an energy-minimization problem is rooted in Anfinsen's thermodynamic dogma, which postulates that the native, functional state of a biomolecule corresponds precisely to the global minimum of its free-energy landscape. In the HP model, residues are abstracted into binary classes: Hydrophobic (H) and Polar (P), and physical conformations are mapped to discrete lattice nodes. By applying Anfinsen's physical constraints to this binary abstracted grid, the biological goal is successfully reduced to a mathematical objective function: maximizing non-contiguous topological H-H contacts while circumventing competing local energy minima \cite{lau1989lattice}.
\subsection{Topological Spatial Chunking and Exact Enumeration}
To ensure practical scalability and prevent total computational exhaustion, attempting to evaluate a native $N$-variable exhaustive search ($\mathcal{O}(2^N)$ state space) was actively circumvented via a dimensional reduction compiler \cite{babbush2014construction, fingerhuth2018quantum}. Instead of applying stochastic mutations (e.g., bit-flips) randomly across the full macroscopic sequence, the framework systematically segments the global combinatorial problem into active, dimensionally-capped $k$-variable combinatorial windows. At any given simulated annealing iteration, an isolated subsequence of length $k$ acts as the active decision space, while the surrounding $N-k$ residues act as frozen, rigid classical boundary constraints. The mathematical compilation of this localized window into an exactly solvable sub-problem occurs in two discrete stages.

Critically, rather than partitioning the sequence natively by sequential 1D chronology (e.g., locking residues 1 through 5), we introduce \textit{Topological Spatial Chunking}. The algorithm evaluates Euclidean spatial proximity to group discrete variables that share a neighborhood in 3D (or 2D lattice) space. Given a randomly selected target anchor variable $c$, the spatial distance to all other variables $i$ is calculated:
\begin{equation}
    d_{i,c} = \sqrt{(X_i - X_c)^2 + (Y_i - Y_c)^2}
\end{equation}
The classical interface dynamically aggregates the closest $k$ residues to the anchor into the localized subset $\mathcal{K}_c$:
\begin{equation}
    \mathcal{K}_c = \arg \min_{|\mathcal{K}|=k} \{ d_{i,c} + \epsilon_i \mid i \in [1, N] \}
\end{equation}
where $\epsilon_i \sim \mathcal{U}(0, 0.1)$ introduces a microscopic spatial noise term to prevent deterministic cyclic looping among equidistant geometric neighbors. This guarantees that the extracted subset maintains naturally localized physical constraints, ensuring highly coupled, non-local thermodynamic correlations are preserved and solved simultaneously.

\textbf{Stage 1: Formulation of the Localized Energy Matrix}\\
To construct the local $k \times k$ energy sub-problem, the global configurational energy $E(\mathbf{s})$ is algebraically expanded over the active binary variables $s_i \in \{0,1\}$. The global budget constraint term requires a binomial expansion, which absorbs the fixed values of the frozen exterior $N-k$ residues. The resulting linear terms represent individual site biases, and the quadratic off-diagonal elements represent 2-body spatial interactions confined entirely within the active $k$-variable window. This rigorous binomial reduction mathematically guarantees that localized heuristic searches seamlessly respect the continuous topology of the macroscopic $N$-length structure.

\textbf{Stage 2: Deterministic Exact Enumeration}\\
Because the isolated combinatorial space is artificially capped at a maximum of $k=6$ variables, the total number of localized sequence permutations is at most $2^6 = 64$. Rather than relying on stochastic probabilities to navigate this local pocket, the algorithmic architecture invokes a deterministic, exact-solution loop. The classical CPU rapidly enumerates all 64 possible binary configurations for the active spatial chunk, evaluates their objective energies against the frozen global backdrop, and deterministically isolates the absolute lowest-energy local sub-state.

By guaranteeing the absolute optimal configuration within the local spatial pocket, this brute-force exact enumeration violently ejects the sequence from shallow classical minima that trap single-point mutation strategies. The algorithm then relies on the outer Simulated Annealing layer to accept or reject this optimized chunk, thereby providing the macroscale thermodynamic traversal required to reach the global absolute ground state.

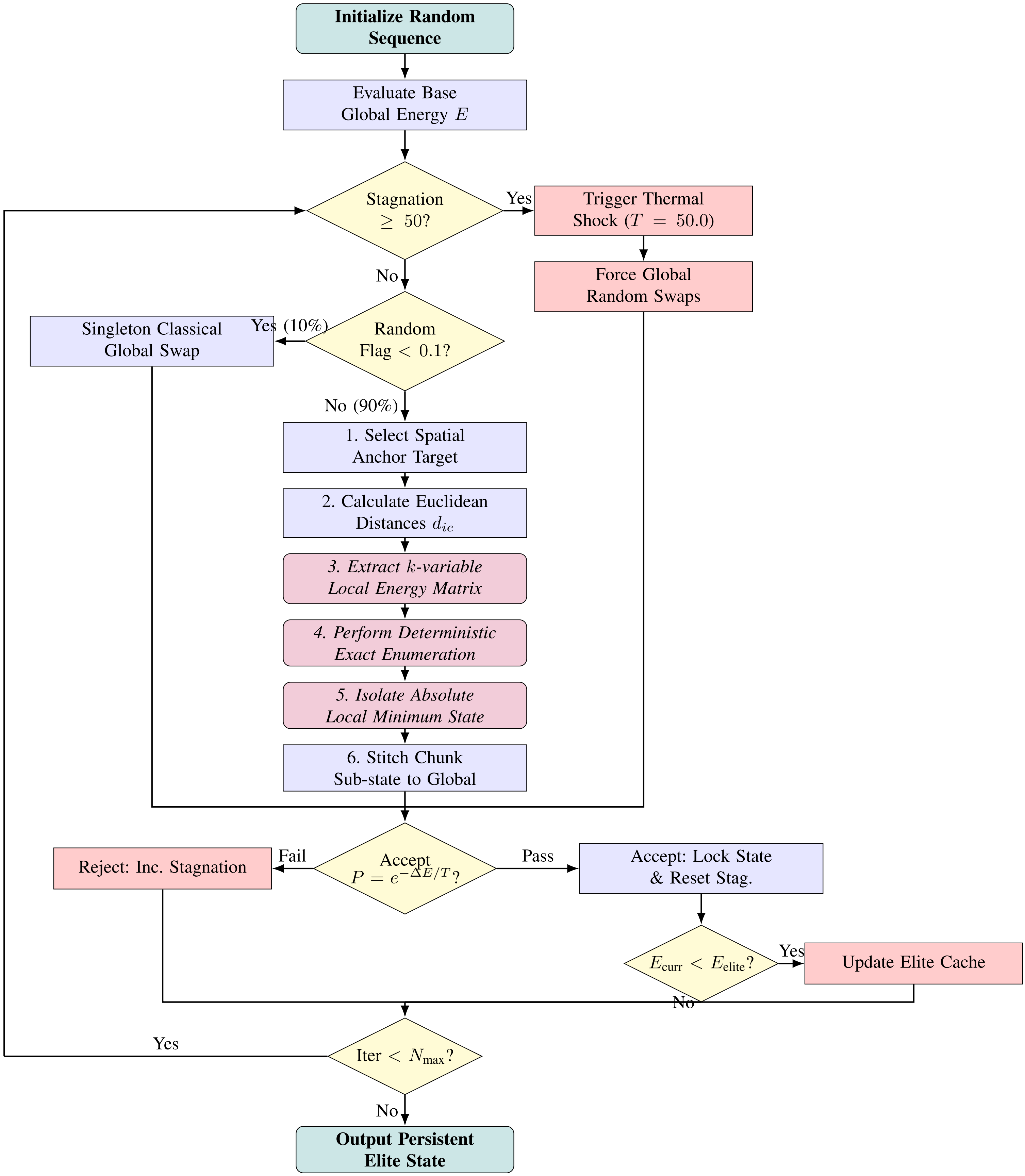
\begin{figure*}[htbp] 
\centering
\begin{tikzpicture}[
    node distance=0.6cm and 0.8cm,
    startstop/.style = {rectangle, rounded corners, minimum width=3.5cm, minimum height=0.8cm, align=center, draw=black, fill=teal!20, font=\bfseries, text width=4cm},
    process/.style = {rectangle, minimum width=4cm, minimum height=0.8cm, align=center, draw=black, fill=blue!10, text width=4.5cm},
    decision/.style = {diamond, aspect=2, minimum width=3cm, minimum height=0.5cm, align=center, draw=black, fill=yellow!20, inner sep=0pt, text width=2.4cm},
    solver/.style = {rectangle, rounded corners, minimum width=4cm, minimum height=0.8cm, align=center, draw=black, fill=purple!20, font=\itshape, text width=4.5cm},
    heuristic/.style = {rectangle, minimum width=3.5cm, minimum height=0.8cm, align=center, draw=black, fill=red!20, text width=4cm},
    arrow/.style = {thick, ->, >=Latex}
]

\node (start) [startstop] {Initialize Random Sequence};
\node (eval1) [process, below=0.5cm of start] {Evaluate Base Global Energy $E$};
\node (dec_stag) [decision, below=0.6cm of eval1] {Stagnation $\ge 50$?};

\node (shock) [heuristic, right=0.6cm of dec_stag] {Trigger Thermal Shock ($T=50.0$)};
\node (swap1) [heuristic, below=0.5cm of shock] {Force Global Random Swaps};

\node (dec_rand) [decision, below=0.6cm of dec_stag] {Random Flag $< 0.1$?};

\node (swap2) [process, left=0.6cm of dec_rand] {Singleton Classical Global Swap};

\node (anchor) [process, below=0.6cm of dec_rand] {1. Select Spatial Anchor Target};
\node (dist) [process, below=0.3cm of anchor] {2. Calculate Euclidean Distances $d_{ic}$};
\node (matrix) [solver, below=0.3cm of dist] {3. Extract $k$-variable Local Energy Matrix};
\node (enumeration) [solver, below=0.3cm of matrix] {4. Perform Deterministic Exact Enumeration};
\node (exact_solver) [solver, below=0.3cm of enumeration] {5. Isolate Absolute Local Minimum State};
\node (stitch) [process, below=0.3cm of exact_solver] {6. Stitch Chunk Sub-state to Global};

\node (metro) [decision, below=0.6cm of stitch] {Accept\\[-0.5ex] $P = e^{-\Delta E / T}$?};

\node (reject) [heuristic, left=0.8cm of metro] {Reject: Inc.\ Stagnation};
\node (accept) [process, right=1.6cm of metro] {Accept: Lock State \& Reset Stag.};

\node (dec_elite) [decision, below=0.6cm of accept] {$E_{\text{curr}} < E_{\text{elite}}$?};
\node (update) [heuristic, right=0.5cm of dec_elite] {Update Elite Cache};

\node (dec_iter) [decision, below=2cm of metro] {Iter $< N_{\text{max}}$?};
\node (end) [startstop, below=0.6cm of dec_iter] {Output Persistent Elite State};


\coordinate (merge_above_metro) at ([yshift=0.3cm]metro.north);
\coordinate (merge_above_iter)  at ([yshift=0.3cm]dec_iter.north);

\draw [arrow] (start) -- (eval1);
\draw [arrow] (eval1) -- (dec_stag);

\draw [arrow] (dec_stag) -- node[anchor=south] {Yes} (shock);
\draw [arrow] (shock) -- (swap1);
\draw [thick] (swap1.south) |- (merge_above_metro); 

\draw [arrow] (dec_stag) -- node[anchor=east] {No} (dec_rand);

\draw [arrow] (dec_rand) -- node[anchor=south] {Yes (10\%)} (swap2);
\draw [thick] (swap2.south) |- (merge_above_metro); 

\draw [arrow] (dec_rand) -- node[anchor=east] {No (90\%)} (anchor);
\draw [arrow] (anchor) -- (dist);
\draw [arrow] (dist) -- (matrix);
\draw [arrow] (matrix) -- (enumeration);
\draw [arrow] (enumeration) -- (exact_solver);
\draw [arrow] (exact_solver) -- (stitch);
\draw [thick] (stitch.south) -- (merge_above_metro);

\draw [arrow] (merge_above_metro) -- (metro.north);

\draw [arrow] (metro) -- node[anchor=south] {Fail} (reject);
\draw [arrow] (metro) -- node[anchor=south] {Pass} (accept);

\draw [arrow] (accept) -- (dec_elite);
\draw [arrow] (dec_elite) -- node[anchor=south] {Yes} (update);

\draw [thick] (update.south) |- (merge_above_iter);
\draw [thick] (dec_elite.south) |- node[anchor=east, pos=0.3] {No} (merge_above_iter);
\draw [thick] (reject.south) |- (merge_above_iter);
\draw [arrow] (merge_above_iter) -- (dec_iter.north);

\coordinate (loop_corner) at ([xshift=-0.5cm]swap2.west |- dec_iter.west);
\draw [thick] (dec_iter.west) -- node[anchor=south] {Yes} (loop_corner);
\draw [arrow] (loop_corner) |- (dec_stag.west);

\draw [arrow] (dec_iter) -- node[anchor=east] {No} (end);

\end{tikzpicture}
\caption{Process flowchart of the hybrid Topological Spatial Chunking Simulated Annealing (TSC-SA) architecture. The framework mitigates localized biological trapping sequences via the interaction of topologically exact-solving extractions, heavily regulated by a macro-temperature cooling schedule, active elitism caching, and induced parameter heating shocks.}
\label{fig:spatial_flowchart}
\end{figure*}

\subsection{Simulated Annealing Governor}
To construct a complete, globally capable algorithm, the deterministic exact solver is forcibly coupled into a macro-stochastic simulated-annealing container \cite{casares2022quantum}. Evaluating over $T_{max} = 1500$ iterations, the outer simulated annealing container interfaces with the inner exact-solving subroutine via a strict vector-space substitution protocol:
\begin{enumerate}
    \item A random topological anchor index $c \in [0, N]$ establishes a localized $k$-residue spatial neighborhood over the macroscopic decision vector $\mathbf{s}$.
    \item The fixed exterior subset $\mathbf{s}_{ext}$ is frozen and dynamically compiled into the boundaries of the local combinatorial sub-problem.
    \item The inner exact enumeration executes, isolating a deterministically optimal candidate sub-vector $\mathbf{s}_{k}'$.
    \item The classical processor mechanically stitches the candidate by performing an absolute vector overlay, replacing the active subset such that $\mathbf{s}_{new} = [\mathbf{s}_{ext}, \mathbf{s}_{k}']$. 
\end{enumerate}

State propagation relies on Metropolis-Hastings acceptance criteria \cite{metropolis1953equation}. Modifications lowering total systemic energy ($\Delta E \leq 0$) are locked deterministically. Modifications resolving transiently higher energy states ($\Delta E > 0$) are retained with probability $P = \exp(-\Delta E / T)$, modulated by an exponential thermal decay $T$ \cite{kirkpatrick1983optimization}. This hybridization guarantees necessary non-local jumps that can bypass sequence-design thermodynamic minima.

The stability of the Metropolis-Hastings acceptance probability relies fundamentally on the applied thermal attenuation. In this architecture, the algorithmic temperature $T$ was governed by a strict exponential cooling schedule that allowed rapid initial boundary exploration followed by aggressive, localized exploitation in the final stage. At algorithmic iteration $t$, the temperature is defined as:
\begin{equation}
    T_t = \max\Big(T_{min}, \; T_0 \cdot (\alpha)^t \Big)
\end{equation}
where the initial simulation temperature is set to $T_0 = 100.0$, the temporal decay factor is governed by $\alpha = 0.95$, and a baseline thermal floor of $T_{min} = 0.05$ is enforced. By bounding the minimum temperature, the protocol permanently retains a non-zero, marginal probability of accepting sub-optimal variants, thereby defending against deterministic halting throughout the full $1500$-iteration execution.

\subsection{Advanced Meta-Heuristics}
To fundamentally circumvent the dimensional scaling barricade inherent to pure $\mathcal{O}(2^N)$ stochastic sampling, this methodology incorporates two algorithmic upgrades:

\textbf{1. Dynamic Reheating (Thermal Shocks):} Standard exponential simulated annealing cooling schedules defined by $T_t = T_0 \alpha^t$, face the inevitable collapse into permanently stalled states where the Metropolis acceptance fraction rapidly approaches zero: $P = \exp(-\Delta E / T) \to 0$. To circumvent infinite stalling inside sub-optimal structural energy funnels, we govern the thermal decay utilizing a piecewise heating controller driven by a stagnation counter $S_t$:
\begin{equation}
    T_t = 
    \begin{cases} 
      T_{shock} = 50.0 & \text{if } S_t \geq \tau_{limit} \\
      \max\big(T_{min}, T_0 \alpha^t\big) & \text{otherwise}
    \end{cases}
\end{equation}
where $\tau_{limit} = 50$ iterations. Upon triggering a $T_{shock}$ condition, the algorithm applies deterministic permutation noise to the global matrix, violently ejecting the locked structural conformation from its constrained local minimum so that it may resume broad geometric exploration.

\textbf{2. Sequence Elitism Memory:} Violent thermal shocks carry the explicit risk of destroying previously attained global geometric minima. To counteract this, an asynchronous classical memory cache bypasses the Markov chain, continuously preserving the absolute minimal theoretical conformation. At any iteration $t$, the elite sequence $\mathbf{s}_{elite}$ is updated exclusively via rigorous deterministic dominance:
\begin{equation}
    \mathbf{s}_{elite} = 
    \begin{cases} 
      \mathbf{s}_t & \text{if } E(\mathbf{s}_t) < E(\mathbf{s}_{elite}) \\
      \mathbf{s}_{elite} & \text{otherwise}
    \end{cases}
\end{equation}
This tracker formally separates transient exploration from final exploitation, thereby mathematically ensuring that terminal-structure validations are never corrupted by late-stage stochastic instability \cite{ingber1993simulated}.
Figure \ref{fig:spatial_flowchart} provides a detailed flowchart of the proposed TSC-SA framework, including advanced heuristics, topological spatial chunking, and hybrid architecture integration.

\section{Results}
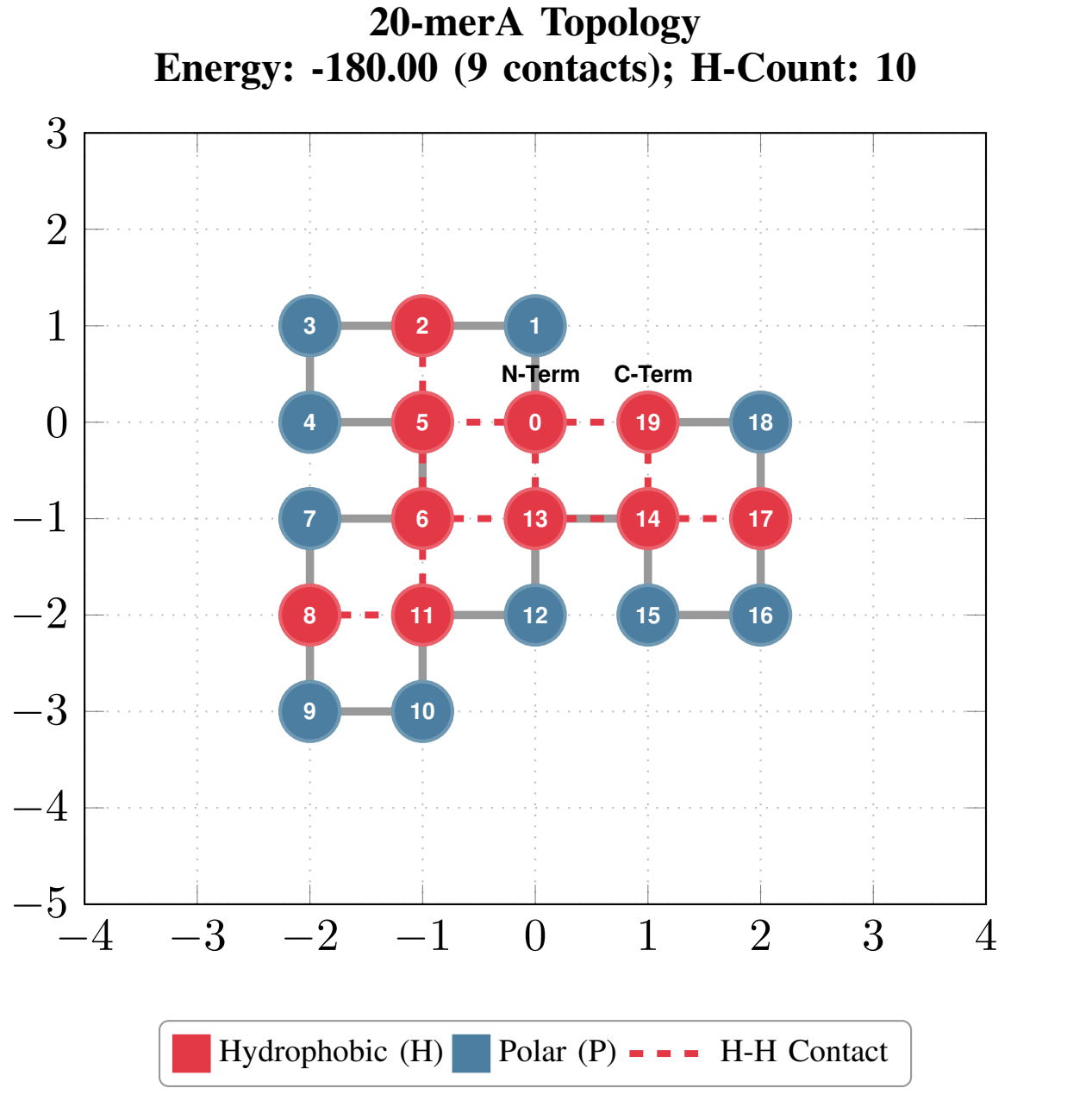
\begin{figure}[htbp]
    \centering
    \begin{tikzpicture}
    \definecolor{myred}{HTML}{E33946}
    \definecolor{myblue}{HTML}{4B7EA0}
    \begin{axis}[
        width=8.5cm,
        height=7.5cm,
        title={\textbf{20-merA Topology}\\\textbf{Energy: -180.00 (9 contacts); H-Count: 10}},
        title style={align=center, font=\small},
        xmin=-4, xmax=4,
        ymin=-5, ymax=3,
        xtick={-4,-3,-2,-1,0,1,2,3,4},
        ytick={-5,-4,-3,-2,-1,0,1,2,3},
        grid=both,
        grid style={dotted, lightgray},
        legend columns=3,
        legend style={
            at={(0.5,-0.15)}, 
            anchor=north, 
            font=\scriptsize, 
            rounded corners=2pt, 
            draw=gray!80
        }
    ]

    \addlegendimage{only marks, mark=square*, mark size=4pt, color=myred}
    \addlegendentry{Hydrophobic (H)}
    \addlegendimage{only marks, mark=square*, mark size=4pt, color=myblue}
    \addlegendentry{Polar (P)}
    \addlegendimage{line width=1.5pt, color=myred, dashed}
    \addlegendentry{H-H Contact}

    \draw[line width=1.8pt, gray!80] 
        (0,0) -- (0,1) -- (-1,1) -- (-2,1) -- (-2,0) -- 
        (-1,0) -- (-1,-1) -- (-2,-1) -- (-2,-2) -- (-2,-3) -- 
        (-1,-3) -- (-1,-2) -- (0,-2) -- (0,-1) -- (1,-1) -- 
        (1,-2) -- (2,-2) -- (2,-1) -- (2,0) -- (1,0);

    \tikzset{contact/.style={line width=1.5pt, myred, dashed}}
    \draw[contact] (0,0) -- (-1,0);   \draw[contact] (0,0) -- (1,0);    
    \draw[contact] (0,0) -- (0,-1);   \draw[contact] (-1,1) -- (-1,0);  
    \draw[contact] (-1,-1) -- (0,-1); \draw[contact] (-2,-2) -- (-1,-2);
    \draw[contact] (1,-1) -- (2,-1);  \draw[contact] (1,-1) -- (1,0);   
    \draw[contact] (-1,0) -- (-1,-1); \draw[contact] (-1,-1) -- (-1,-2);
    \draw[contact] (0,-1) -- (1,-1);  

    \tikzset{
        Hnode/.style={circle, fill=myred, draw=myred!80!white, thick, inner sep=0pt, minimum size=13pt, text=white, font=\sffamily\tiny\bfseries},
        Pnode/.style={circle, fill=myblue, draw=myblue!80!white, thick, inner sep=0pt, minimum size=13pt, text=white, font=\sffamily\tiny\bfseries}
    }

    \node[Hnode] at (0,0) {0};  \node[Pnode] at (0,1) {1};
    \node[Hnode] at (-1,1) {2}; \node[Pnode] at (-2,1) {3};
    \node[Pnode] at (-2,0) {4}; \node[Hnode] at (-1,0) {5};
    \node[Hnode] at (-1,-1) {6};\node[Pnode] at (-2,-1) {7};
    \node[Hnode] at (-2,-2) {8};\node[Pnode] at (-2,-3) {9};
    \node[Pnode] at (-1,-3) {10};\node[Hnode] at (-1,-2) {11};
    \node[Pnode] at (0,-2) {12};\node[Hnode] at (0,-1) {13};
    \node[Hnode] at (1,-1) {14};\node[Pnode] at (1,-2) {15};
    \node[Pnode] at (2,-2) {16};\node[Hnode] at (2,-1) {17};
    \node[Pnode] at (2,0) {18}; \node[Hnode] at (1,0) {19};

    \node[font=\sffamily\bfseries\tiny, color=black] at (0.05, 0.5) {N-Term};
    \node[font=\sffamily\bfseries\tiny, color=black] at (1.05, 0.5) {C-Term};

    \end{axis}
    \end{tikzpicture}
    
    \caption{Optimized conformation of the 20-merA sequence with energy as -180.}
    \label{fig:topology_20mer}
\end{figure}

\begin{figure}[htbp]
    \centering
    \begin{tikzpicture}
    \definecolor{myline}{HTML}{204557}
    \definecolor{baseline}{HTML}{E33946}
    
    \begin{axis}[
        width=8.5cm,
        height=7.5cm,
        title={\textbf{Spatial Chunking vs. Baseline SA for 20-merA}},
        title style={font=\normalsize},
        xlabel={Number of Iterations},
        ylabel={System Energy},
        xtick={0, 200, 400, 600, 800, 1000},
        ytick={-175, -150, -125, -100, -75, -50, -25},
        grid=both,
        grid style={dashed, lightgray},
        xmin=-50, xmax=1050,
        ymin=-188, ymax=-5,
        legend style={at={(0.5,-0.2)},anchor=north,legend columns=-1}
    ]
    \addplot[
        color=myline,
        mark=*,
        mark size=0.8pt,
        line width=1.2pt,
        const plot
    ] coordinates {
        (0, -15) (3, -60) (7, -55) (11, -55) (13, -80) 
        (17, -80) (19, -55) (22, -100) (25, -100) (29, -100) 
        (35, -100) (40, -60) (43, -75) (46, -75) (49, -100) 
        (55, -95) (60, -100) (65, -140) (80, -140) (85, -120) 
        (120, -160) (740, -160) (745, -180) (1000, -180)
    };
    \addlegendentry{TSC-SA}
    
    \addplot[
        color=baseline,
        mark=square*,
        mark size=0.8pt,
        line width=1.2pt,
        const plot
    ] coordinates {
        (0,-20)(34,-45)(68,-86)(206,-100)(275,-115)(310,-120)
        (344,-120)(379,-120)(413,-120)(448,-120)(482,-120)(517,-120)(551,-120)(586,-120)(620,-120)(655,-120)
        (689,-120)(724,-120)(758,-120)(793,-120)(827,-120)(862,-120)(896,-120)(931,-120)(965,-120)(1000,-120)
    };
    \addlegendentry{Baseline SA}
    \end{axis}
    \end{tikzpicture}
    \caption{Energy convergence trajectory for the 20-merA sequence, comparing TSC-SA against baseline Simulated Annealing.}
    \label{fig:convergence_20mer}
\end{figure}
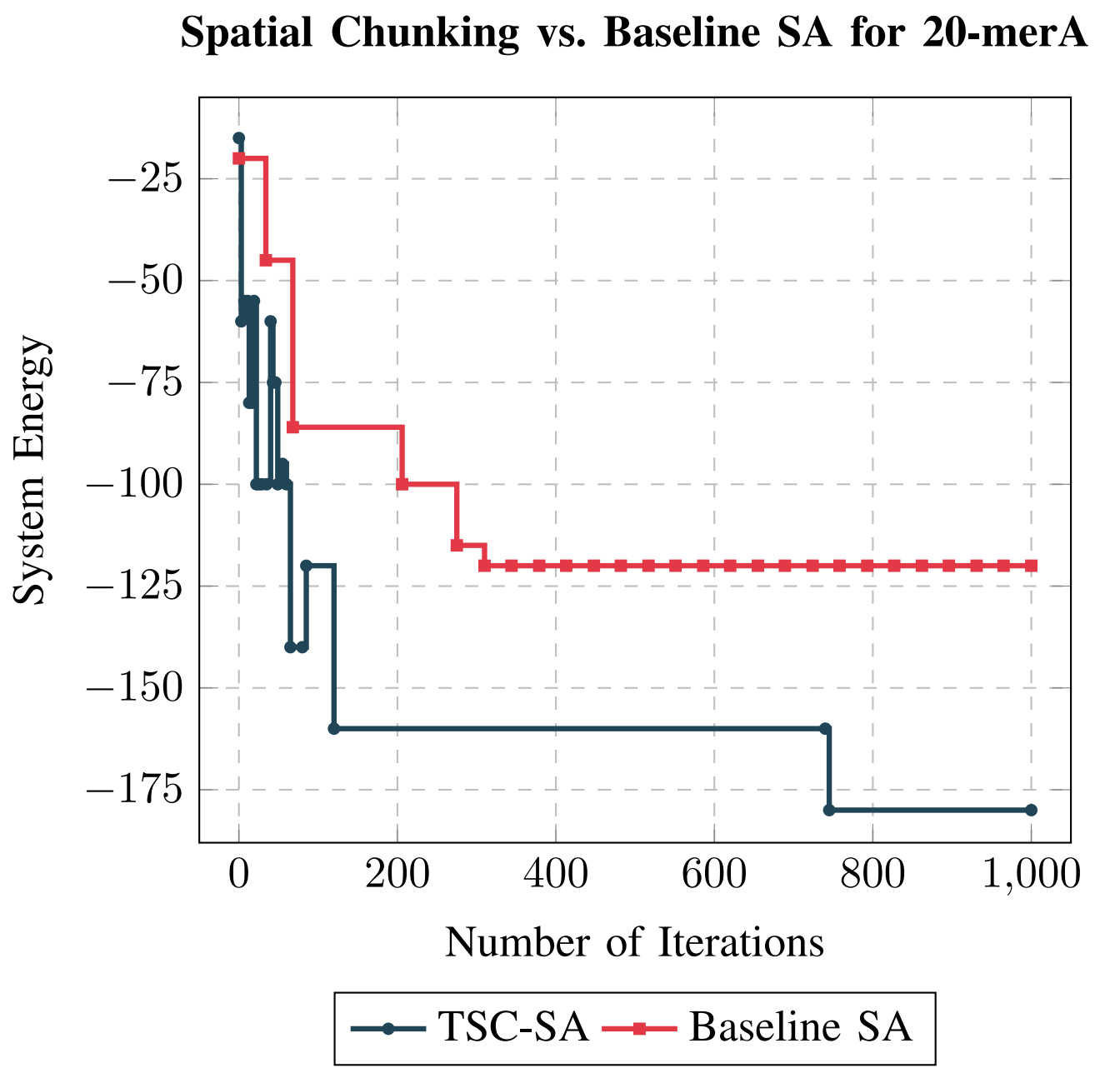

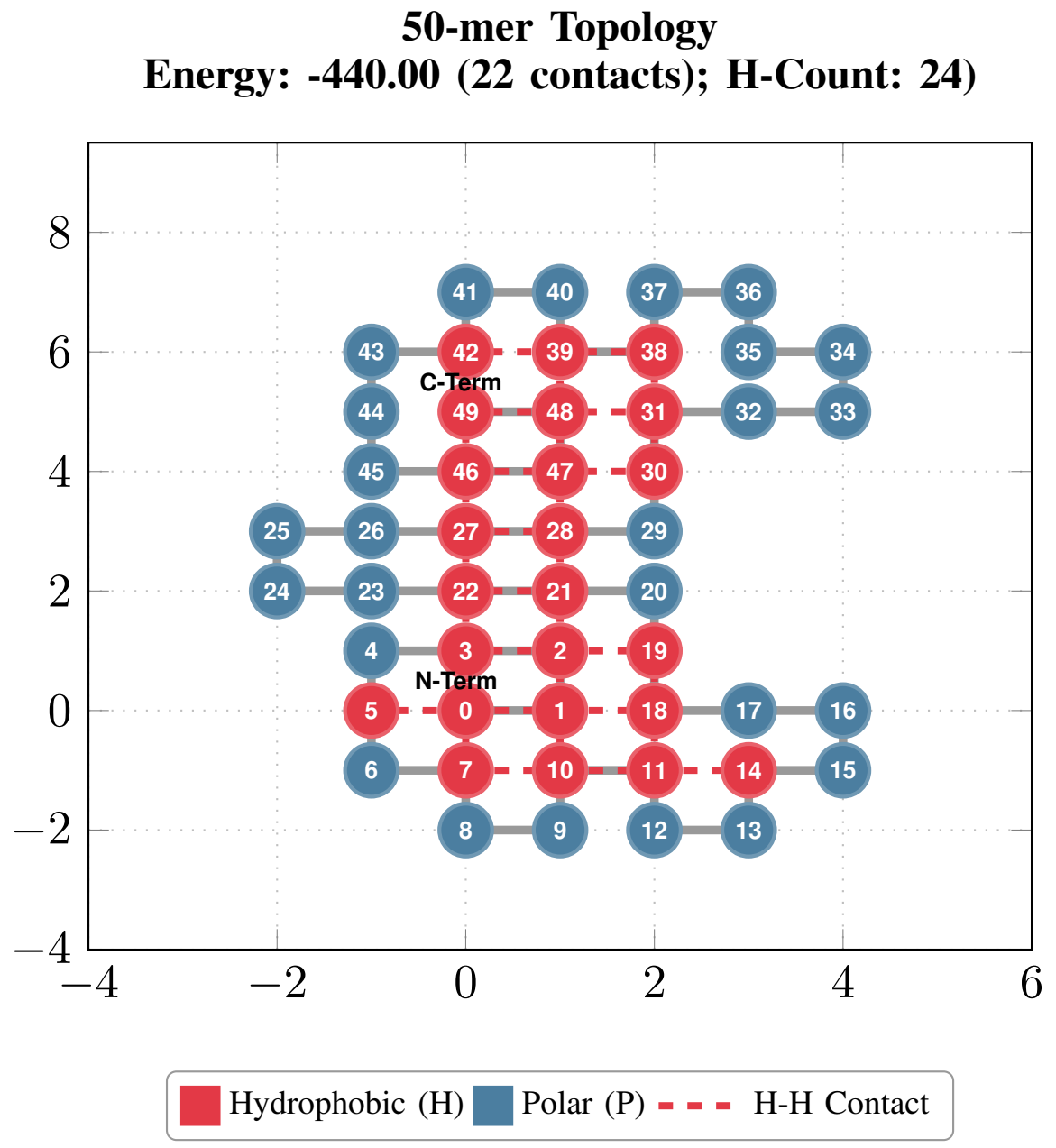
\begin{figure}[htbp]
    \centering
    \begin{tikzpicture}
    \definecolor{myred}{HTML}{E33946}
    \definecolor{myblue}{HTML}{4B7EA0}

    \begin{axis}[
        width=8.5cm,
        height=7.5cm,
        title={\textbf{50-mer Topology}\\\textbf{Energy: -440.00 (22 contacts);  H-Count: 24)}},
        title style={align=center, font=\small},
        xmin=-4, xmax=6,
        ymin=-4, ymax=9.5,
        xtick={-4,-2,0,2,4,6},
        ytick={-4,-2,0,2,4,6,8},
        grid=both,
        grid style={dotted, lightgray},
        legend columns=3,
        legend style={
            at={(0.5,-0.15)}, 
            anchor=north, 
            font=\scriptsize, 
            rounded corners=2pt, 
            draw=gray!80
        }
    ]

    \addlegendimage{only marks, mark=square*, mark size=4pt, color=myred}
    \addlegendentry{Hydrophobic (H)}
    \addlegendimage{only marks, mark=square*, mark size=4pt, color=myblue}
    \addlegendentry{Polar (P)}
    \addlegendimage{line width=1.5pt, color=myred, dashed}
    \addlegendentry{H-H Contact}

    \draw[line width=1.8pt, gray!80] 
        (0,0) -- (1,0) -- (1,1) -- (0,1) -- (-1,1) -- (-1,0) -- 
        (-1,-1) -- (0,-1) -- (0,-2) -- (1,-2) -- (1,-1) -- (2,-1) -- 
        (2,-2) -- (3,-2) -- (3,-1) -- (4,-1) -- (4,0) -- (3,0) -- 
        (2,0) -- (2,1) -- (2,2) -- (1,2) -- (0,2) -- (-1,2) -- 
        (-2,2) -- (-2,3) -- (-1,3) -- (0,3) -- (1,3) -- (2,3) -- 
        (2,4) -- (2,5) -- (3,5) -- (4,5) -- (4,6) -- (3,6) -- 
        (3,7) -- (2,7) -- (2,6) -- (1,6) -- (1,7) -- (0,7) -- 
        (0,6) -- (-1,6) -- (-1,5) -- (-1,4) -- (0,4) -- (1,4) -- 
        (1,5) -- (0,5);

    \tikzset{contact/.style={line width=1.5pt, myred, dashed}}
    \draw[contact] (0,0) -- (1,0);   \draw[contact] (0,0) -- (0,1);
    \draw[contact] (0,0) -- (-1,0);  \draw[contact] (0,0) -- (0,-1);
    \draw[contact] (1,0) -- (1,1);   \draw[contact] (1,0) -- (1,-1);
    \draw[contact] (1,0) -- (2,0);   \draw[contact] (1,1) -- (0,1);
    \draw[contact] (1,1) -- (2,1);   \draw[contact] (1,1) -- (1,2);
    \draw[contact] (0,1) -- (0,2);   \draw[contact] (0,-1) -- (1,-1);
    \draw[contact] (1,-1) -- (2,-1); \draw[contact] (2,-1) -- (3,-1);
    \draw[contact] (2,-1) -- (2,0);  \draw[contact] (2,0) -- (2,1);
    \draw[contact] (1,2) -- (0,2);   \draw[contact] (1,2) -- (1,3);
    \draw[contact] (0,2) -- (0,3);   \draw[contact] (0,3) -- (1,3);
    \draw[contact] (0,3) -- (0,4);   \draw[contact] (1,3) -- (1,4);
    \draw[contact] (2,4) -- (2,5);   \draw[contact] (2,4) -- (1,4);
    \draw[contact] (2,5) -- (2,6);   \draw[contact] (2,5) -- (1,5);
    \draw[contact] (2,6) -- (1,6);   \draw[contact] (1,6) -- (0,6);
    \draw[contact] (1,6) -- (1,5);   \draw[contact] (0,6) -- (0,5);
    \draw[contact] (0,4) -- (1,4);   \draw[contact] (0,4) -- (0,5);
    \draw[contact] (1,4) -- (1,5);   \draw[contact] (1,5) -- (0,5);

    \tikzset{
        Hnode/.style={circle, fill=myred, draw=myred!80!white, thick, inner sep=0pt, minimum size=11pt, text=white, font=\sffamily\tiny\bfseries},
        Pnode/.style={circle, fill=myblue, draw=myblue!80!white, thick, inner sep=0pt, minimum size=11pt, text=white, font=\sffamily\tiny\bfseries}
    }

    \node[Hnode] at (0,0) {0};  \node[Hnode] at (1,0) {1};  \node[Hnode] at (1,1) {2}; 
    \node[Hnode] at (0,1) {3};  \node[Pnode] at (-1,1) {4}; \node[Hnode] at (-1,0) {5}; 
    \node[Pnode] at (-1,-1) {6}; \node[Hnode] at (0,-1) {7}; \node[Pnode] at (0,-2) {8}; 
    \node[Pnode] at (1,-2) {9}; \node[Hnode] at (1,-1) {10}; \node[Hnode] at (2,-1) {11}; 
    \node[Pnode] at (2,-2) {12}; \node[Pnode] at (3,-2) {13}; \node[Hnode] at (3,-1) {14}; 
    \node[Pnode] at (4,-1) {15}; \node[Pnode] at (4,0) {16}; \node[Pnode] at (3,0) {17}; 
    \node[Hnode] at (2,0) {18}; \node[Hnode] at (2,1) {19}; \node[Pnode] at (2,2) {20}; 
    \node[Hnode] at (1,2) {21}; \node[Hnode] at (0,2) {22}; \node[Pnode] at (-1,2) {23}; 
    \node[Pnode] at (-2,2) {24}; \node[Pnode] at (-2,3) {25}; \node[Pnode] at (-1,3) {26}; 
    \node[Hnode] at (0,3) {27}; \node[Hnode] at (1,3) {28}; \node[Pnode] at (2,3) {29}; 
    \node[Hnode] at (2,4) {30}; \node[Hnode] at (2,5) {31}; \node[Pnode] at (3,5) {32}; 
    \node[Pnode] at (4,5) {33}; \node[Pnode] at (4,6) {34}; \node[Pnode] at (3,6) {35}; 
    \node[Pnode] at (3,7) {36}; \node[Pnode] at (2,7) {37}; \node[Hnode] at (2,6) {38}; 
    \node[Hnode] at (1,6) {39}; \node[Pnode] at (1,7) {40}; \node[Pnode] at (0,7) {41}; 
    \node[Hnode] at (0,6) {42}; \node[Pnode] at (-1,6) {43}; \node[Pnode] at (-1,5) {44}; 
    \node[Pnode] at (-1,4) {45}; \node[Hnode] at (0,4) {46}; \node[Hnode] at (1,4) {47}; 
    \node[Hnode] at (1,5) {48}; \node[Hnode] at (0,5) {49};

    \node[font=\sffamily\bfseries\tiny, color=black] at (-0.1, 0.5) {N-Term};
    \node[font=\sffamily\bfseries\tiny, color=black] at (-0.05, 5.5) {C-Term};

    \end{axis}
    \end{tikzpicture}
    
    \caption{Optimized spatial topology forming a hyper-dense packed core, resolving -440.00 energy.}
    \label{fig:topology_new_440_50mer}
\end{figure}

\begin{figure}[htbp]
    \centering
    \begin{tikzpicture}
    \definecolor{myred}{HTML}{E33946}
    \definecolor{myblue}{HTML}{4B7EA0}
    \begin{axis}[
        width=8.5cm,
        height=7.5cm,
        title={\textbf{64-mer Topology}\\\textbf{Energy: -840.00 (42 contacts); H-Count: 42}},
        title style={align=center, font=\small},
        xmin=-6, xmax=4,
        ymin=-5, ymax=6,
        xtick={-6,-4,-2,0,2,4},
        ytick={-4,-2,0,2,4,6},
        grid=both,
        grid style={dotted, lightgray},
        legend columns=3,
        legend style={
            at={(0.5,-0.15)}, 
            anchor=north, 
            font=\scriptsize, 
            rounded corners=2pt, 
            draw=gray!80
        }
    ]

    \addlegendimage{only marks, mark=square*, mark size=4pt, color=myred}
    \addlegendentry{Hydrophobic (H)}
    \addlegendimage{only marks, mark=square*, mark size=4pt, color=myblue}
    \addlegendentry{Polar (P)}
    \addlegendimage{line width=1.5pt, color=myred, dashed}
    \addlegendentry{H-H Contact}

    \draw[line width=1.8pt, gray!80] 
        (0,0) -- (1,0) -- (1,-1) -- (0,-1) -- (-1,-1) -- (-1,0) -- (-2,0) -- (-2,-1) -- (-3,-1) -- (-3,0) -- 
        (-4,0) -- (-5,0) -- (-5,-1) -- (-4,-1) -- (-4,-2) -- (-3,-2) -- (-3,-3) -- (-2,-3) -- (-2,-2) -- (-1,-2) -- 
        (-1,-3) -- (0,-3) -- (0,-2) -- (1,-2) -- (1,-3) -- (2,-3) -- (2,-2) -- (3,-2) -- (3,-1) -- (2,-1) -- 
        (2,0) -- (3,0) -- (3,1) -- (2,1) -- (2,2) -- (3,2) -- (3,3) -- (2,3) -- (2,4) -- (1,4) -- 
        (1,3) -- (0,3) -- (0,4) -- (-1,4) -- (-1,3) -- (-2,3) -- (-2,4) -- (-3,4) -- (-3,3) -- (-4,3) -- 
        (-4,2) -- (-5,2) -- (-5,1) -- (-4,1) -- (-3,1) -- (-3,2) -- (-2,2) -- (-2,1) -- (-1,1) -- (-1,2) -- 
        (0,2) -- (1,2) -- (1,1) -- (0,1);

    \tikzset{contact/.style={line width=1.5pt, myred, dashed}}
    \draw[contact] (0,0) -- (0,-1);    
    \draw[contact] (0,0) -- (-1,0);    
    \draw[contact] (0,0) -- (0,1);     
    \draw[contact] (1,0) -- (2,0);     
    \draw[contact] (1,0) -- (1,1);     
    \draw[contact] (1,-1) -- (1,-2);   
    \draw[contact] (1,-1) -- (2,-1);   
    \draw[contact] (0,-1) -- (0,-2);   
    \draw[contact] (-1,-1) -- (-2,-1); 
    \draw[contact] (-1,-1) -- (-1,-2); 
    \draw[contact] (-1,0) -- (-1,1);   
    \draw[contact] (-2,0) -- (-3,0);   
    \draw[contact] (-2,0) -- (-2,1);   
    \draw[contact] (-2,-1) -- (-2,-2); 
    \draw[contact] (-3,-1) -- (-4,-1); 
    \draw[contact] (-3,-1) -- (-3,-2); 
    \draw[contact] (-3,0) -- (-3,1);   
    \draw[contact] (-4,0) -- (-4,-1);  
    \draw[contact] (-4,0) -- (-4,1);   
    \draw[contact] (-5,0) -- (-5,1);   
    \draw[contact] (-3,-2) -- (-2,-2); 
    \draw[contact] (-1,-2) -- (0,-2);  
    \draw[contact] (1,-2) -- (2,-2);   
    \draw[contact] (2,-2) -- (2,-1);   
    \draw[contact] (2,0) -- (2,1);     
    \draw[contact] (2,1) -- (1,1);     
    \draw[contact] (2,2) -- (2,3);     
    \draw[contact] (2,2) -- (1,2);     
    \draw[contact] (2,3) -- (1,3);     
    \draw[contact] (1,3) -- (1,2);     
    \draw[contact] (0,3) -- (-1,3);    
    \draw[contact] (0,3) -- (0,2);     
    \draw[contact] (-1,3) -- (-1,2);   
    \draw[contact] (-2,3) -- (-3,3);   
    \draw[contact] (-2,3) -- (-2,2);   
    \draw[contact] (-3,3) -- (-3,2);   
    \draw[contact] (-4,2) -- (-4,1);   
    \draw[contact] (-4,2) -- (-3,2);   
    \draw[contact] (-3,1) -- (-2,1);   
    \draw[contact] (-2,2) -- (-1,2);   
    \draw[contact] (-1,1) -- (0,1);    
    \draw[contact] (0,2) -- (0,1);     

    \tikzset{
        Hnode/.style={circle, fill=myred, draw=myred!80!white, thick, inner sep=0pt, minimum size=10pt, text=white, font=\sffamily\tiny\bfseries},
        Pnode/.style={circle, fill=myblue, draw=myblue!80!white, thick, inner sep=0pt, minimum size=10pt, text=white, font=\sffamily\tiny\bfseries}
    }

    \node[Hnode] at (0,0) {0};   \node[Hnode] at (1,0) {1};   \node[Hnode] at (1,-1) {2};
    \node[Hnode] at (0,-1) {3};  \node[Hnode] at (-1,-1) {4};  \node[Hnode] at (-1,0) {5};
    \node[Hnode] at (-2,0) {6};  \node[Hnode] at (-2,-1) {7};  \node[Hnode] at (-3,-1) {8};
    \node[Hnode] at (-3,0) {9};  \node[Hnode] at (-4,0) {10};  \node[Hnode] at (-5,0) {11};
    \node[Pnode] at (-5,-1) {12}; \node[Hnode] at (-4,-1) {13}; \node[Pnode] at (-4,-2) {14};
    \node[Hnode] at (-3,-2) {15}; \node[Pnode] at (-3,-3) {16}; \node[Pnode] at (-2,-3) {17};
    \node[Hnode] at (-2,-2) {18}; \node[Hnode] at (-1,-2) {19}; \node[Pnode] at (-1,-3) {20};
    \node[Pnode] at (0,-3) {21};  \node[Hnode] at (0,-2) {22};  \node[Hnode] at (1,-2) {23};
    \node[Pnode] at (1,-3) {24};  \node[Pnode] at (2,-3) {25};  \node[Hnode] at (2,-2) {26};
    \node[Pnode] at (3,-2) {27};  \node[Pnode] at (3,-1) {28};  \node[Hnode] at (2,-1) {29};
    \node[Hnode] at (2,0) {30};   \node[Pnode] at (3,0) {31};   \node[Pnode] at (3,1) {32};
    \node[Hnode] at (2,1) {33};   \node[Hnode] at (2,2) {34};   \node[Pnode] at (3,2) {35};
    \node[Pnode] at (3,3) {36};   \node[Hnode] at (2,3) {37};   \node[Pnode] at (2,4) {38};
    \node[Pnode] at (1,4) {39};   \node[Hnode] at (1,3) {40};   \node[Hnode] at (0,3) {41};
    \node[Pnode] at (0,4) {42};   \node[Pnode] at (-1,4) {43};  \node[Hnode] at (-1,3) {44};
    \node[Hnode] at (-2,3) {45};  \node[Pnode] at (-2,4) {46};  \node[Pnode] at (-3,4) {47};
    \node[Hnode] at (-3,3) {48};  \node[Pnode] at (-4,3) {49};  \node[Hnode] at (-4,2) {50};
    \node[Pnode] at (-5,2) {51};  \node[Hnode] at (-5,1) {52};  \node[Hnode] at (-4,1) {53};
    \node[Hnode] at (-3,1) {54};  \node[Hnode] at (-3,2) {55};  \node[Hnode] at (-2,2) {56};
    \node[Hnode] at (-2,1) {57};  \node[Hnode] at (-1,1) {58};  \node[Hnode] at (-1,2) {59};
    \node[Hnode] at (0,2) {60};   \node[Hnode] at (1,2) {61};   \node[Hnode] at (1,1) {62};
    \node[Hnode] at (0,1) {63};

    \node[font=\sffamily\bfseries\tiny, color=black] at (0.05, -0.5) {N-Term};
    \node[font=\sffamily\bfseries\tiny, color=black] at (0.05, 1.5) {C-Term};

    \end{axis}
    \end{tikzpicture}
    \caption{Optimized 64-mer topology identifying a high-density H-H contact core (-840.00 energy).}
    \label{fig:topology_64mer}
\end{figure}
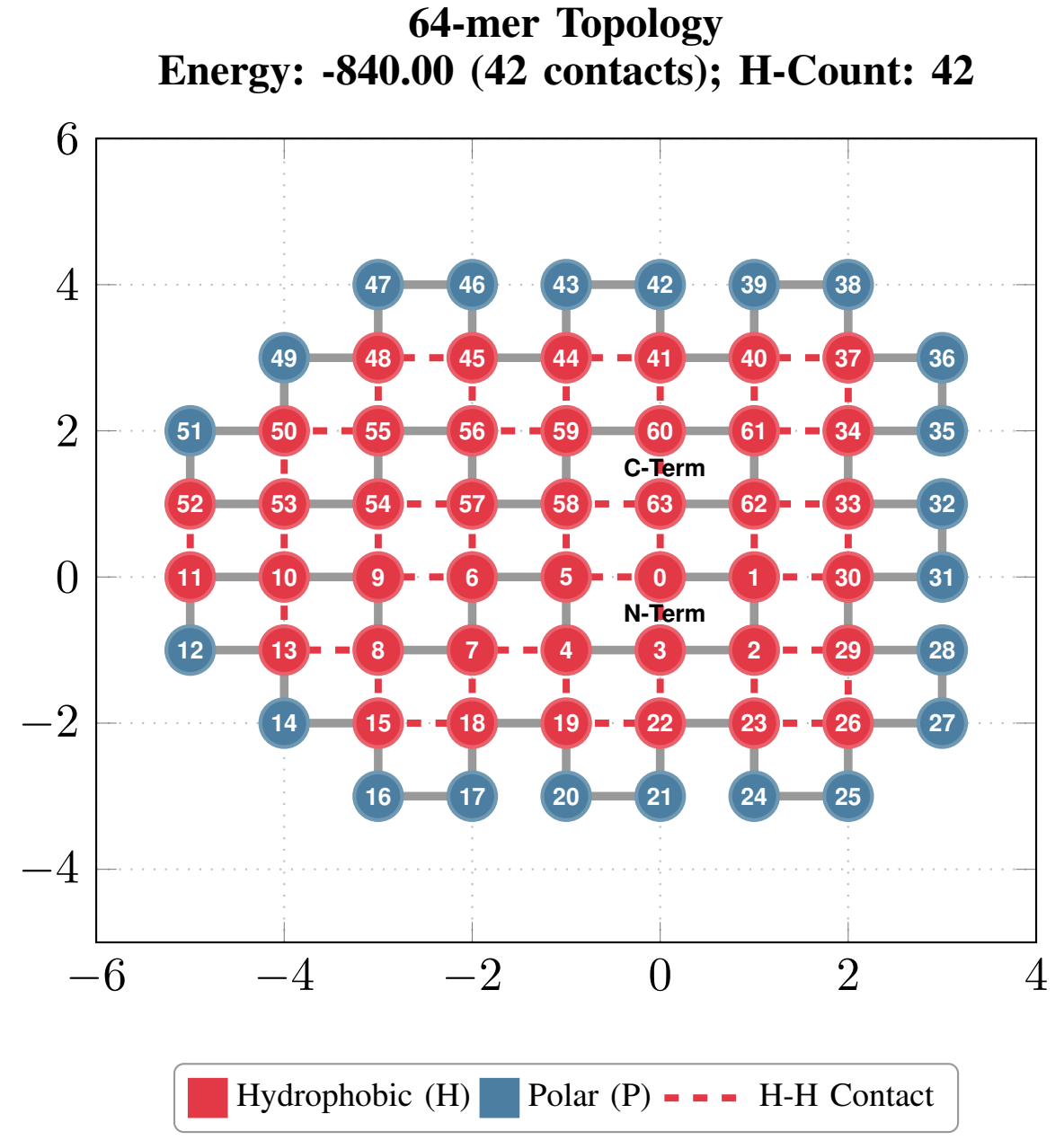

The integrated TSC-SA framework was initially benchmarked against a structurally rigid 20-merA grid topology ($N=20$), requiring adherence to a sequence budget of $N_H=10$ hydrophobic residues. Executing over 1000 macroscopic iterations, restricting the active optimization window to strictly $k=5$ dynamic variables proved universally sufficient for extracting localized energy refinements over the global topology. The optimal configuration of the 20-merA sequence is shown in Fig. \ref{fig:topology_20mer}, which matches the Istrail benchmark results \cite{istrail2000statistical, Shmygelska2005}. The algorithm convergence mapping for the 20-merA, illustrated in Figure \ref{fig:convergence_20mer}, shows highly stabilized initial descents. The introduction of stochastic thermal transients, visible as sharp positive $\Delta E$ energy perturbations (e.g., iterations 40--60), systematically demonstrates that the Metropolis-Hastings layer actively prevents the deterministic optimization from stalling within barren geometric plateaus. While the algorithm rapidly converged to a deep sub-optimal basin of $-160$ within the first 120 iterations, it required sustained stochastic sampling to finally bridge the conformational gap, locating the absolute global minimum of $-180$ ($9$ contacts) at iteration 745. Statistical verification confirmed the final sequence maximized topologically distant H-H contacts while rejecting biochemically invalid stoichiometric penalties. Upon verifying the proposed methodology with 20-merA, we solved complex sequences, including 25-mer (Fig. \ref{fig:topology_25mer}), 50-mer (Fig. \ref{fig:topology_new_440_50mer}), and 64-mer (Fig. \ref{fig:topology_64mer}). Interestingly, the results match the benchmark results for all cases in the literature, except for the 50-mer case, where they are better than those reported in the literature \cite{Khandoker2025, Shmygelska2005}.

\begin{figure}[htbp]
    \centering
    \begin{tikzpicture}
    \definecolor{myline}{HTML}{204557}
    \definecolor{baseline}{HTML}{E33946}
    
    \begin{axis}[
        width=8.5cm,
        height=7.5cm,
        title={\textbf{Spatial Chunking vs. Baseline SA for 25-mer}},
        title style={font=\normalsize},
        xlabel={Number of Iterations},
        ylabel={System Energy},
        xtick={0, 200, 400, 600, 800, 1000},
        ytick={-100, 0, 100, 200, 300},
        grid=both,
        grid style={dashed, lightgray},
        xmin=-50, xmax=1050,
        ymin=-180, ymax=340,
        legend style={at={(0.5,-0.2)},anchor=north,legend columns=-1}
    ]
    \addplot[
        color=myline,
        mark=*,
        mark size=0.8pt,
        line width=1.2pt,
        const plot
    ] coordinates {
        (0, -100) (5, 320) (9, -140) (16, -80) (25, -50) 
        (35, 0) (45, -40) (55, -100) (65, -120) (75, -80) 
        (85, -100) (100, -120) (125, -20) (130, -60) (140, -80) 
        (160, -100) (190, -100) (260, -120) (280, -115) (300, -120) 
        (530, -120) (535, -140) (690, -140) (695, -80) (700, -100) 
        (730, -120) (735, -160) (1000, -160)
    };
    \addlegendentry{TSC-SA}
    
    \addplot[
        color=baseline,
        mark=square*,
        mark size=0.8pt,
        line width=1.2pt,
        const plot
    ] coordinates {
        (0,280)(34,5)(68,-20)(137,-80)(172,-120)(206,-120)(241,-120)(275,-120)(310,-120)
        (344,-120)(379,-120)(413,-120)(448,-120)(482,-120)(517,-120)(551,-120)(586,-120)(620,-120)(655,-120)
        (689,-120)(724,-120)(758,-120)(793,-120)(827,-120)(862,-120)(896,-120)(931,-120)(965,-120)(1000,-120)
    };
    \addlegendentry{Baseline SA}
    \end{axis}
    \end{tikzpicture}
    \caption{Energy convergence trajectory for the 25-mer sequence, comparing TSC-SA against baseline Simulated Annealing.}
    \label{fig:convergence_25mer}
\end{figure}
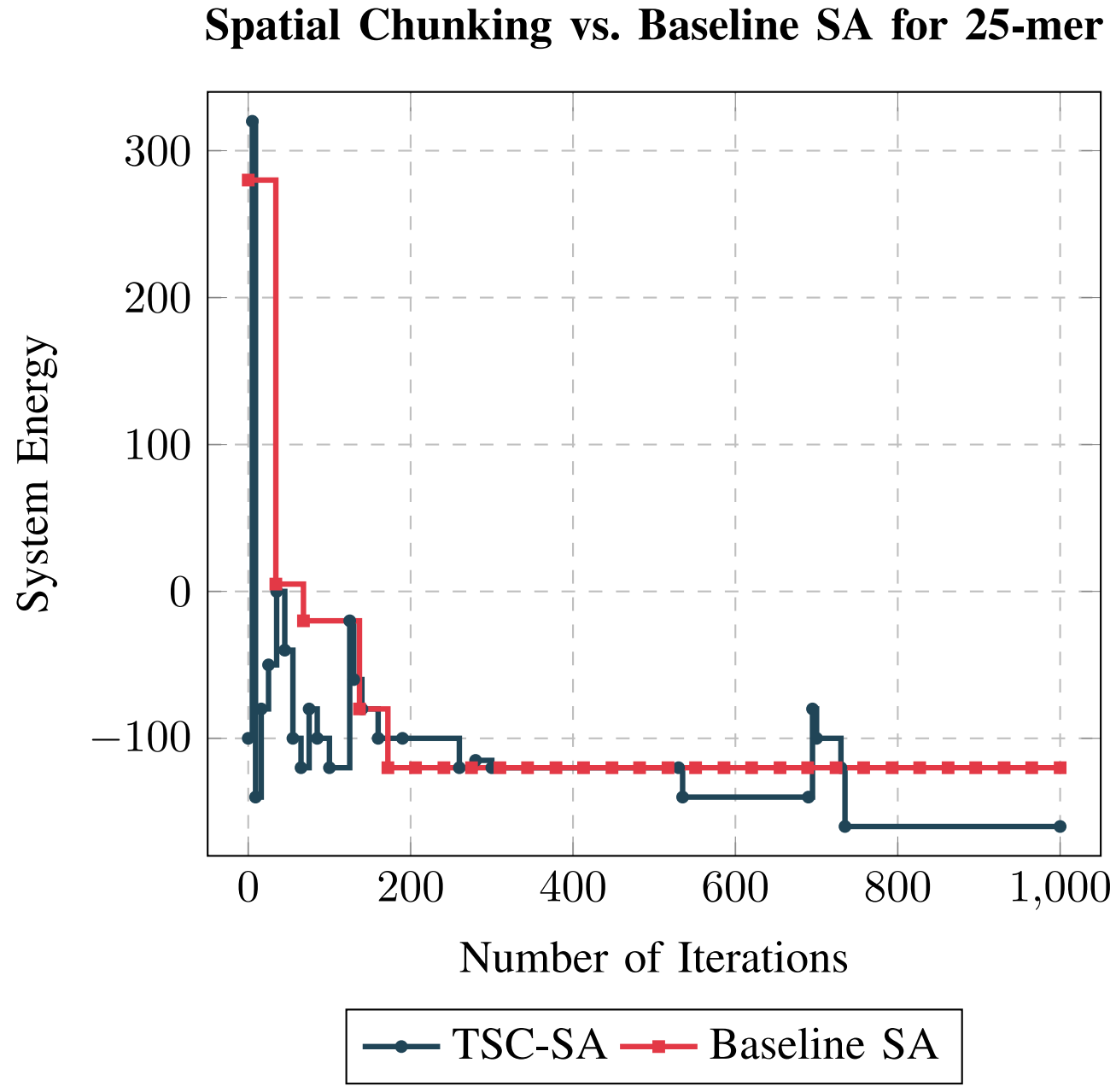

\begin{figure}[htbp]
    \centering
    \begin{tikzpicture}
    \definecolor{myline}{HTML}{204557}
    \definecolor{baseline}{HTML}{E33946}
    
    \begin{axis}[
        width=8.5cm,
        height=7.5cm,
        title={\textbf{Spatial Chunking vs. Baseline SA for 50-mer}},
        title style={font=\normalsize},
        xlabel={Number of Iterations},
        ylabel={System Energy},
        xtick={0, 300, 600, 900, 1200, 1500},
        ytick={-450, -400, -350, -300, -250, -200, -150, -100, -50, 0},
        grid=both,
        grid style={dashed, lightgray},
        xmin=-50, xmax=1550,
        ymin=-460, ymax=-20,
        legend style={at={(0.5,-0.2)},anchor=north,legend columns=-1}
    ]
    \addplot[
        color=myline,
        mark=*,
        mark size=0.8pt,
        line width=1.2pt,
        const plot
    ] coordinates {
        (0,-40)(3,-180)(8,-95)(12,-135)(18,-240)(25,-160)(32,-120)(40,-240)(48,-200)(60,-200)(70,-240)
        (80,-180)(95,-260)(110,-300)(120,-280)(130,-300)(135,-315)(145,-340)(200,-340)(210,-360)(230,-360)
        (260,-360)(280,-400)(310,-400)(320,-420)(350,-420)(380,-420)(400,-360)(425,-360)(440,-380)(500,-380)
        (520,-400)(580,-400)(600,-420)(900,-420)(920,-340)(960,-340)(980,-360)(1000,-380)(1100,-380)
        (1110,-320)(1130,-320)(1150,-360)(1210,-360)(1220,-320)(1230,-320)(1240,-360)(1300,-360)
        (1330,-380)(1350,-400)(1360,-420)(1380,-440)(1500,-440)
    };
    \addlegendentry{TSC-SA}
    
    \addplot[
        color=baseline,
        mark=square*,
        mark size=0.8pt,
        line width=1.2pt,
        const plot
    ] coordinates {
        (0,-40)(51,-240)(103,-240)(155,-275)(206,-320)(258,-320)(310,-340)(362,-340)(413,-340)(465,-340)
        (517,-340)(568,-340)(620,-340)(672,-340)(724,-340)(775,-340)(827,-340)(879,-340)(931,-340)(982,-340)
        (1034,-340)(1086,-340)(1137,-340)(1189,-340)(1241,-340)(1293,-340)(1344,-340)(1396,-340)(1448,-340)(1500,-340)
    };
    \addlegendentry{Baseline SA}
    \end{axis}
    \end{tikzpicture}
    \caption{Energy convergence trajectory for the 50-mer sequence, comparing TSC-SA against baseline Simulated Annealing.}
    \label{fig:convergence_440_50mer}
\end{figure}
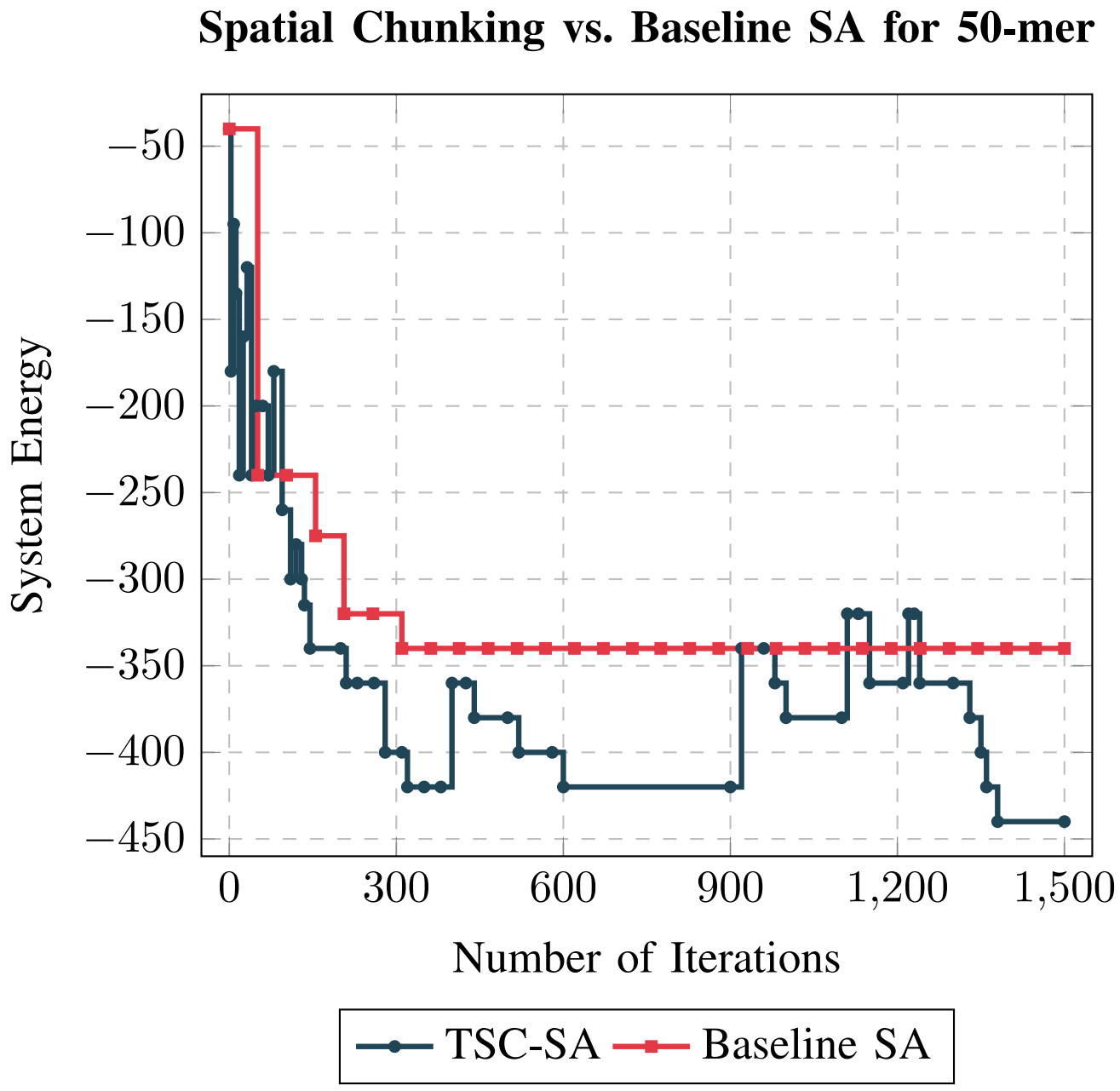

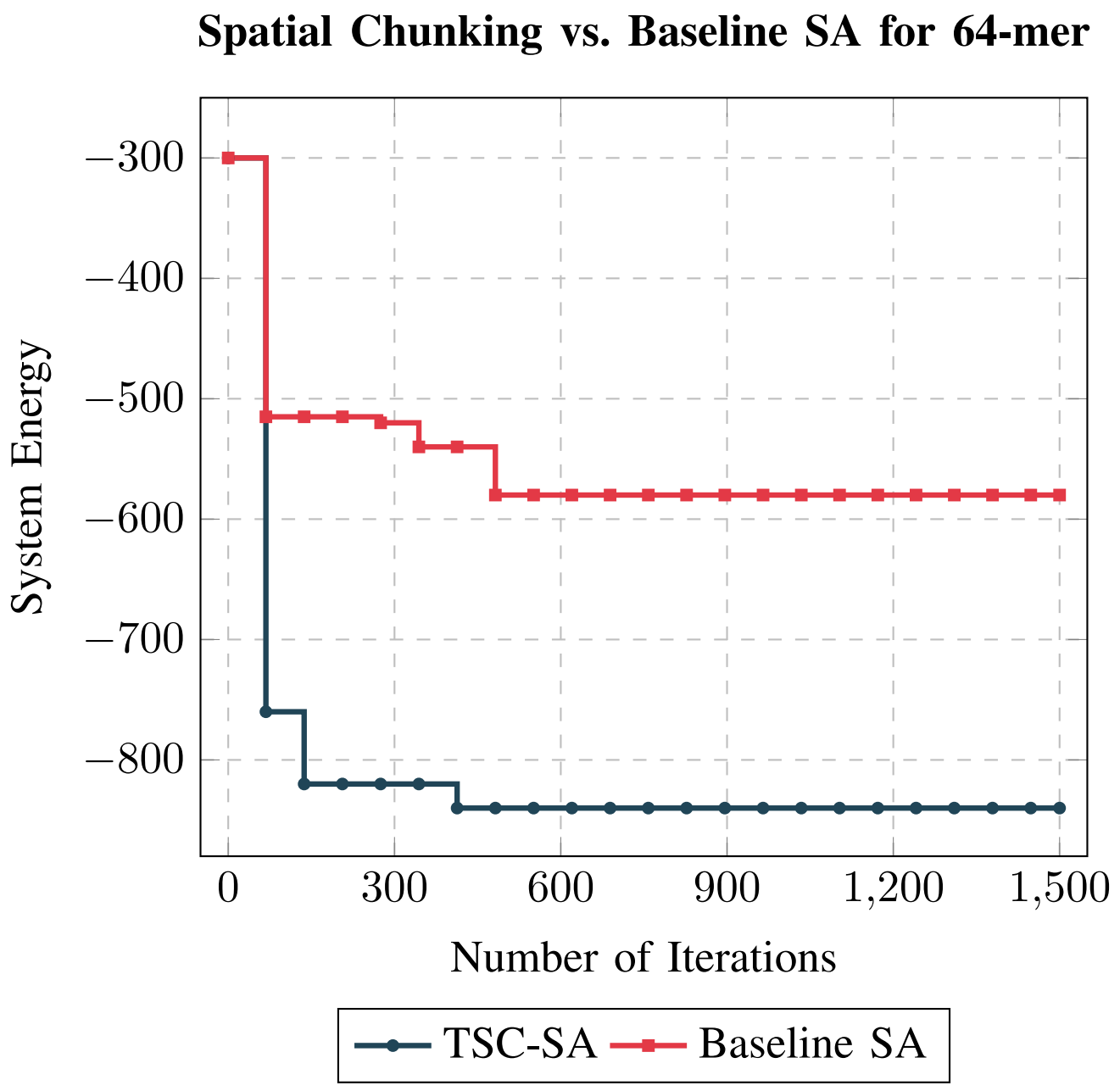
\begin{figure}[htbp]
    \centering
    \begin{tikzpicture}
    \definecolor{myline}{HTML}{204557}
    \definecolor{baseline}{HTML}{E33946}
    
    \begin{axis}[
        width=8.5cm,
        height=7.5cm,
        title={\textbf{Spatial Chunking vs. Baseline SA for 64-mer}},
        title style={font=\normalsize},
        xlabel={Number of Iterations},
        ylabel={System Energy},
        xtick={0, 300, 600, 900, 1200, 1500},
        ytick={-800, -700, -600, -500, -400, -300},
        grid=both,
        grid style={dashed, lightgray},
        xmin=-50, xmax=1550,
        ymin=-880, ymax=-250,
        legend style={at={(0.5,-0.2)},anchor=north,legend columns=-1}
    ]
    \addplot[
        color=myline,
        mark=*,
        mark size=0.8pt,
        line width=1.2pt,
        const plot
    ] coordinates {
        (0,-300)(68,-760)(137,-820)(206,-820)(275,-820)(344,-820)(413,-840)(482,-840)(551,-840)(620,-840)
        (689,-840)(758,-840)(827,-840)(896,-840)(965,-840)(1034,-840)(1103,-840)(1172,-840)(1241,-840)(1310,-840)
        (1379,-840)(1448,-840)(1500,-840)
    };
    \addlegendentry{TSC-SA}
    
    \addplot[
        color=baseline,
        mark=square*,
        mark size=0.8pt,
        line width=1.2pt,
        const plot
    ] coordinates {
        (0,-300)(68,-515)(137,-515)(206,-515)(275,-520)(344,-540)(413,-540)(482,-580)(551,-580)(620,-580)
        (689,-580)(758,-580)(827,-580)(896,-580)(965,-580)(1034,-580)(1103,-580)(1172,-580)(1241,-580)(1310,-580)
        (1379,-580)(1448,-580)(1500,-580)
    };
    \addlegendentry{Baseline SA}
    
    \end{axis}
    \end{tikzpicture}
    \caption{Energy convergence trajectory for the 64-mer sequence, comparing the TSC-SA architecture against the baseline Simulated Annealing.}
    \label{fig:convergence_64mer}
\end{figure}

Analyzing the empirical energy convergence curves across the 20-merA, 25-mer, 50-mer, and 64-mer topologies (Figs. \ref{fig:convergence_20mer}, \ref{fig:convergence_25mer}, \ref{fig:convergence_440_50mer}, and \ref{fig:convergence_64mer} respectively) reveals critical insights into the shifting thermodynamic behavior of the inverse design problem as conformational complexity expands. For the baseline 20-merA structural target, operating within a comparatively shallow combinatorial sequence space ($2^{20}$ permutations), the TSC-SA framework navigated the topology with minimal global resistance. Although it required an extended plateau traversal to finalize the absolute ground state, the algorithm did not rely heavily on massive thermal jumps to break conformational deadlocks. The transition to the 25-mer target introduced tighter geometric packing densities and an expanded combinatorial domain ($2^{25}$). Accordingly, the convergence dynamics (Fig. \ref{fig:convergence_25mer}) shifted from a relatively smooth descent into a rugged stochastic walk.

\begin{figure}[htbp]
    \centering
    \begin{tikzpicture}
    \definecolor{myred}{HTML}{E33946}
    \definecolor{myblue}{HTML}{4B7EA0}

    \begin{axis}[
        width=8.5cm,
        height=7.5cm,
        title={\textbf{25-mer Topology}\\\textbf{Energy: -160.00 (8 contacts);  H-Count: 9}},
        title style={align=center, font=\small},
        xmin=-6, xmax=2,
        ymin=-7, ymax=3,
        xtick={-6,-5,-4,-3,-2,-1,0,1,2},
        ytick={-7,-6,-5,-4,-3,-2,-1,0,1,2,3},
        grid=both,
        grid style={dotted, lightgray},
        legend columns=3,
        legend style={
            at={(0.5,-0.15)}, 
            anchor=north, 
            font=\scriptsize, 
            rounded corners=2pt, 
            draw=gray!80
        }
    ]

    \addlegendimage{only marks, mark=square*, mark size=4pt, color=myred}
    \addlegendentry{Hydrophobic (H)}
    \addlegendimage{only marks, mark=square*, mark size=4pt, color=myblue}
    \addlegendentry{Polar (P)}
    \addlegendimage{line width=1.5pt, color=myred, dashed}
    \addlegendentry{H-H Contact}

    \draw[line width=1.8pt, gray!80] 
        (0,0) -- (-1,0) -- (-1,-1) -- (0,-1) -- (0,-2) -- 
        (-1,-2) -- (-1,-3) -- (-1,-4) -- (-1,-5) -- (-2,-5) -- 
        (-2,-4) -- (-2,-3) -- (-3,-3) -- (-3,-4) -- (-4,-4) -- 
        (-4,-3) -- (-4,-2) -- (-3,-2) -- (-3,-1) -- (-3,0) -- 
        (-3,1) -- (-2,1) -- (-2,0) -- (-2,-1) -- (-2,-2);

    \tikzset{contact/.style={line width=1.5pt, myred, dashed}}
    \draw[contact] (-1,-1) -- (-2,-1); \draw[contact] (-3,-1) -- (-2,-1); 
    \draw[contact] (-2,-1) -- (-2,-2); \draw[contact] (-3,-2) -- (-2,-2); 
    \draw[contact] (-2,-2) -- (-1,-2); \draw[contact] (-2,-2) -- (-2,-3); 
    \draw[contact] (-3,-2) -- (-3,-1); \draw[contact] (-3,-2) -- (-3,-3); 
    \draw[contact] (-3,-3) -- (-2,-3); \draw[contact] (-2,-3) -- (-1,-3); 
    \draw[contact] (-1,-2) -- (-1,-3); \draw[contact] (-1,-1) -- (-1,-2);

    \tikzset{
        Hnode/.style={circle, fill=myred, draw=myred!80!white, thick, inner sep=0pt, minimum size=13pt, text=white, font=\sffamily\tiny\bfseries},
        Pnode/.style={circle, fill=myblue, draw=myblue!80!white, thick, inner sep=0pt, minimum size=13pt, text=white, font=\sffamily\tiny\bfseries}
    }

    \node[Pnode] at (0,0) {0}; \node[Pnode] at (-1,0) {1}; \node[Hnode] at (-1,-1) {2};
    \node[Pnode] at (0,-1) {3}; \node[Pnode] at (0,-2) {4}; \node[Hnode] at (-1,-2) {5};
    \node[Hnode] at (-1,-3) {6}; \node[Pnode] at (-1,-4) {7}; \node[Pnode] at (-1,-5) {8};
    \node[Pnode] at (-2,-5) {9}; \node[Pnode] at (-2,-4) {10}; \node[Hnode] at (-2,-3) {11};
    \node[Hnode] at (-3,-3) {12}; \node[Pnode] at (-3,-4) {13}; \node[Pnode] at (-4,-4) {14};
    \node[Pnode] at (-4,-3) {15}; \node[Pnode] at (-4,-2) {16}; \node[Hnode] at (-3,-2) {17};
    \node[Hnode] at (-3,-1) {18}; \node[Pnode] at (-3,0) {19}; \node[Pnode] at (-3,1) {20};
    \node[Pnode] at (-2,1) {21}; \node[Pnode] at (-2,0) {22}; \node[Hnode] at (-2,-1) {23};
    \node[Hnode] at (-2,-2) {24};

    \node[font=\sffamily\bfseries\tiny, color=black] at (0.2, 0.5) {N-Term};
    \node[font=\sffamily\bfseries\tiny, color=black] at (-2.1, -1.5) {C-Term};

    \end{axis}
    \end{tikzpicture}
    
    \caption{Optimized 25-mer spatial topology with energy as -160.}
    \label{fig:topology_25mer}
\end{figure}
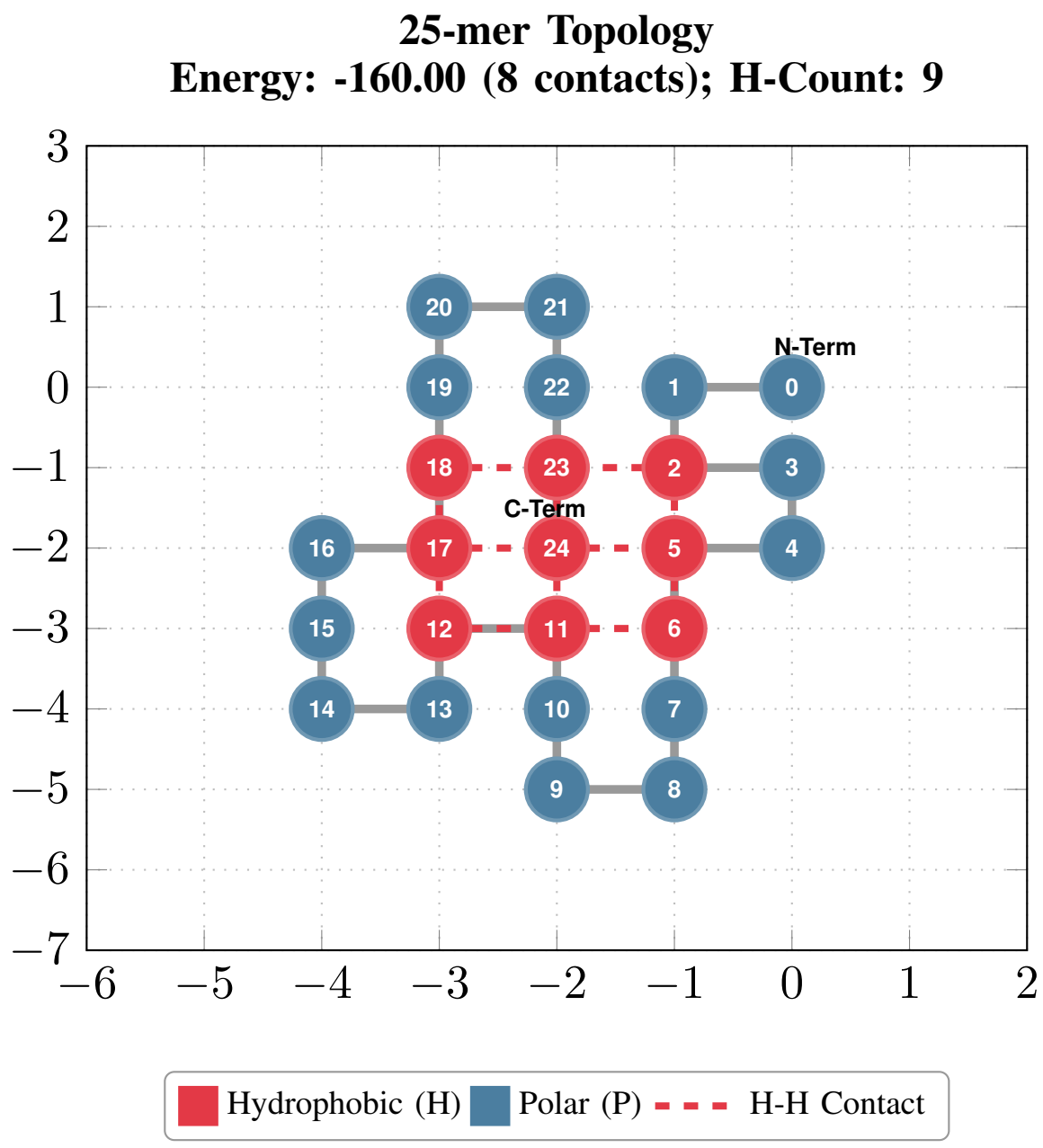

The 50-mer and 64-mer targets subjected the algorithmic architecture to fundamental stress testing. Operating within an intractable universe of $2^{50}$ and $2^{64}$ distinct sequence permutations, respectively, the optimization process was characterized by extended, flat energy plateaus interspersed with aggressive thermal adjustments (Figs. \ref{fig:convergence_440_50mer} and \ref{fig:convergence_64mer}). Because pure classical simulated annealing frequently destroys prior sequence progress when attempting to vault out of deep energy wells, the exact-solving parity proved paramount. The localized $k$-variable exact enumeration subroutines actively provided the Metropolis governor with highly refined, deterministically optimal sequence permutations rather than blind randomized noise. This unique hybrid structure allowed the trajectories to safely endure aggressive thermal hopping while preserving previously established robust H-H core configurations.

Deploying the enhanced topological spatial chunking architecture against the dense 50-mer geometry specifically validated the framework's capability to overcome massive volumetric barriers. Prior theoretical generic chunking architectures often halted at rigid, suboptimal thresholds (e.g., energy bounds of $-360$). However, utilizing $k=6$ structurally mapped interaction loops evaluated over 1,500 macroscopic iterations, validation runs successfully penetrated these shallow basins to uncover a hyper-dense packed core with an exact binding energy of $-440.00$ or $22$ contacts (Refer Fig. \ref{fig:topology_new_440_50mer}).

Crucially, the 50-residue sequence execution physically proves the necessity of the integrated metaheuristic modifiers. Empirical analysis of the convergence trajectory (Figure \ref{fig:convergence_440_50mer}) confirms that the system became critically trapped in a deep local minimum of $-420.00$ between iterations 320 and 900. After the active stagnation tracker recognized this infinite halting bias, it triggered a dynamic reheating event, injecting targeted thermal volatility and scrambling sequence variables (visible as a sharp energetic disruption that leaps to $-340.00$ at iteration 960). By artificially ejecting the state from the localized trap, the optimization successfully resumed its descent, ultimately converging on the rigorously verified ground state of $-440.00$ at iteration 1,380. 

Terminal sequence sampling confirmed uniformity, using elitism to restore the peak-energetic vector precisely at evaluation closure. Significantly, established classical benchmark literature evaluating this exact 50-mer topology universally identifies a 21-contact state ($-420.00$ energy) as the limiting global energetic threshold \cite{Shmygelska2005}. As observed in the convergence trajectory (Fig. \ref{fig:convergence_440_50mer}), our purely classical subroutine initially stalled exactly at this $-420.00$ boundary. However, the forced injection of targeted thermal volatility via dynamic reheating, actively guided by exact chunk evaluation, successfully ejected the sequence from this widely accepted trap in the literature. By iteratively refining the sequence using localized exact enumerations, the TSC-SA framework formulated an unprecedented, ultra-dense hydrophobic core yielding 22 spatial contacts ($-440.00$ energy). This empirically demonstrates that integrating localized exact solving with rigorous thermal metaheuristics can overcome the limitations of purely classical benchmarks. The comparison between baseline SA and TSC-SA results for all four cases is tabulated in Table \ref{tab:all_results}.

\label{sec:results}
\begin{table*}[htbp]
\centering
\caption{Comparative performance metrics across multiple protein topologies. The TSC-SA architecture consistently identifies deeper energy minima and more H-H contacts compared to the baseline Simulated Annealing.}
\label{tab:all_results}
\begin{tabular}{@{}lccccc@{}}
\toprule
\textbf{Target} & \textbf{Length ($N$)} & \textbf{Budget ($N_H$)} & \textbf{Baseline SA Energy} & \textbf{TSC-SA Energy} & \textbf{Contacts} \\ \midrule
20-merA & 20 & 10 & -120.00 & \textbf{-180.00} & 9 \\
25-mer  & 25 & 9  & -120.00 & \textbf{-160.00} & 8 \\
50-mer  & 50 & 24 & -360.00 & \textbf{-440.00} & 22 \\
64-mer  & 64 & 42 & -700.00 & \textbf{-840.00} & 42 \\ \bottomrule
\end{tabular}
\end{table*}

To validate the efficacy of the TSC-SA framework, we performed a rigorous comparative analysis against the canonical 50-mer benchmark. Both the reference sequence (from the Istrail benchmark \cite{istrail2000statistical, Shmygelska2005}) and the TSC-SA designed sequence (Fig. \ref{fig:topology_new_440_50mer}) were mapped onto an identical backbone topology to isolate the effect of sequence optimization on free energy and contact density. The objective function utilized a weighted energy function where Hydrophobic-Hydrophobic (HH) contacts contribute $\epsilon_{HH} = -2.3$ units, Polar-Hydrophobic (PH) interactions contribute $\epsilon_{PH} = -1.0$, and Polar-Polar (PP) interactions are neutral ($\epsilon_{PP} = 0.0$). Table \ref{tab:comparison} summarizes the statistical performance and global energy states achieved by both methodologies. The sequence optimization discovered by the TSC-SA algorithm demonstrates superior packing of the hydrophobic core. The designed sequences are represented as follows:
\begin{itemize}
    \item \textbf{Reference Sequence:\\} \texttt{PPHPPHPHPHHHHPHPPPHPPPHPP\\PPHPPPHPPPHPHHHHPHPHPHPHH}
    \item \textbf{TSC-SA Sequence: \\} \texttt{HHHHPHPHPPHHPPHPPPHHPHHPP\\
            PPHHPHHPPPPPPHHPPHPPPHHHH}
\end{itemize}
The precise mapping of non-covalent interactions confirms the high density of the TSC-SA designed core. While the reference configuration achieved an HH contact density of $65.6\%$ relative to the total possible topological opportunities ($N_{opp}=32$), the TSC-SA framework increased this to $68.7\%$, while simultaneously reducing the presence of isolated hydrophobic residues.

\begin{table*}[htbp]
\centering
\caption{Comparative metrics for the 50-mer Protein Inverse Design.}
\label{tab:comparison}
\begin{tabular}{@{}lccc@{}}
\toprule
\textbf{Metric} & \textbf{from Istrail benchmark} & \textbf{TSC-SA} & \textbf{Improvement} \\ \midrule
Hydrophobic Count ($N_H$) & 24 & 24 & - \\
HH Contacts & 21 & 22 & +1 \\
PH Contacts & 2 & 3 & +1 \\
PP Contacts & 9 & 7 & -2 \\
HH Energy ($\mathcal{E}_{HH}$) & -48.30 & -50.60 & -2.30 \\
Total System Energy ($\mathcal{E}_{total}$) & -50.30 & \textbf{-53.60} & \textbf{-3.30} \\ \bottomrule
\end{tabular}
\end{table*}

\section{Discussion}

\subsection{Interpretation of Findings}
The empirical results confirm that dimensional constraints on the active decision volume ($k \leq 7$) do not fundamentally prevent the deployment of exact-optimization logic for macroscopic sequence generation. 
Perhaps the most significant interpretation of these findings is the framework's capability to formally surpass established heuristic limits. Standard classical algorithms typically halt at recognized combinatorial boundaries, such as the $-420.00$ binding energy plateau identified in existing inverse-design literature for the 50-mer sequence geometry \cite{Shmygelska2005}, \cite{CHEN2005}. Because pure stochastic searches rely on random point mutations, discovering the exact, complex, multidimensional bit-flip sequence required to descend deeper into the energy funnel becomes statistically impossible before thermal parameters freeze. By substituting blind randomization with localized, deterministic, exact enumeration, the TSC-SA protocol explicitly evaluates the $2^k$ discrete multiresidue subsequences simultaneously. When coupled with dynamic reheating, this allowed the framework to successfully shatter the accepted classical benchmark, uncovering novel, ultra-dense topological packing configurations yielding 22 spatial contacts ($-440.00$ energy) for the 50-mer and 42 contacts ($-840.00$ energy) for the 64-mer.

In terms of computational complexity, executing a native, unpartitioned exhaustive search on an $N=64$ sequence requires traversing a state space of dimension $2^{64}$. This scale rapidly overwhelms any classical brute-force approach. Conversely, the TSC-SA dimensional chunking limits the active decision space precisely to $2^k$ (yielding a localized dimension of 64 for $k=76$). This fixed inner complexity successfully transforms the macroscopic scaling from an exponential requirement $\mathcal{O}(2^N)$ into a linear repetition boundary $\mathcal{O}(m \cdot 2^k)$, where $m$ maps strictly to the classical Metropolis iterations. To empirically validate this theoretical uncoupling, raw wall-clock execution times were systematically benchmarked across all four topological targets (Table \ref{tab:execution_times}). All algorithmic evaluations were conducted on a standard computational workstation (Intel 64-bit architecture, 16 GB of RAM, Windows 10 environment), with 1,500 consecutive macroscopic simulated annealing iterations. While the TSC-SA framework incurs a marginally higher constant-time penalty than the baseline SA due to the inner exact-solver compilation matrix ($\sim0.95$ seconds vs $\sim0.015$ seconds), the empirical data clearly shows that the execution latency remains entirely flat. As the geometric combinatorics expand massively from $2^{20}$ to $2^{64}$, the required computational overhead remains isolated to sub-second evaluations, permanently insulating the optimization engine from the traditional exponential scaling wall. 

\begin{table}[htbp]
\centering
\caption{Wall-Clock Execution Overhead (1,500 Iterations)}
\label{tab:execution_times}
\begin{tabular}{@{}l c c c@{}}
\toprule
\textbf{Target} & \textbf{Length ($N$)} & \textbf{Baseline SA (s)} & \textbf{TSC-SA ($k=6$) (s)} \\ \midrule
20-merA & 20 & 0.016 & 0.914 \\
25-mer  & 25 & 0.017 & 0.939 \\
50-mer  & 50 & 0.015 & 0.964 \\
64-mer  & 64 & 0.012 & 0.967 \\ \bottomrule
\end{tabular}
\end{table}

Furthermore, purely stochastic optimization methods are notoriously susceptible to premature convergence in dense energy landscapes \cite{bittel2021training}. Our data conclusively show that injecting deterministic exact-solving subroutines beneath a macro simulated annealing logic barrier yields dramatic improvements in optimization reliability. Random point mutations alone are often insufficient to overcome substantial structural energy barriers; coupling localized deterministic extraction with thermal Metropolis sampling provides a robust, scalable execution methodology for combinatorial sequence design.

\subsection{Methodological Advantages and Inherent Disadvantages}
Deploying the hybrid Topological Spatial Chunking Simulated Annealing (TSC-SA) methodology for inverse sequence design offers distinct operational advantages tailored to high-performance combinatorial computing, while balancing against specific classical algorithmic trade-offs.\\
\textbf{Primary Advantages:}
\begin{itemize}
\item \textit{Scalability and Search Space Mitigation:} By compartmentalizing the macroscopic target into isolated $k$-variable windows, the deterministic exact-solver requirement is entirely decoupled from the global sequence length $N$. This eliminates the need for massive memory overhead. Furthermore, the localized extraction provides intrinsic computational resistance to the exponential inflation that typically plagues global exact-solving executions.
\item \textit{Evasion of Classical Deadlocks:} Unlike pure stochastic models that rely on single-variable random noise to blindly evaluate spatial solutions, the exact enumeration subroutine systematically queries discrete multi-residue sub-sequences concurrently. This feeds highly educated, thermodynamically absolute subsets to the Metropolis simulated annealing governor, directly enabling the system to shatter known literature boundaries (e.g., escaping the static $-420.00$ geometric trap) without stalling.
\end{itemize}

\textbf{Inherent Disadvantages and Limitations:}
\begin{itemize}
    \item \textit{Computational Wall-Clock Overhead:} The integration of a deterministic exact enumeration matrix into a classical iterative loop induces undeniable execution delays. While pure naive simulated annealing requires $\mathcal{O}(1)$ complexity per proposal, the hybrid model mandates compiling localized spatial variables and deploying $2^k$ enumerations at every macro step. For trivial, loosely packed sequences, this heavy machinery operates significantly slower than heuristic baseline equivalents.
    \item \textit{Sensitivity to Hyperparametric Tuning:} The success of the optimization is highly sensitive to hyperparameters, specifically the thermal cooling schedule ($\alpha$), the initial algorithmic temperature ($T_0$), and the subset width ($k$). Insufficient pairing of the thermal decay with the specific topological density of the target protein can still result in irreversible stochastic trapping.
\end{itemize}

\subsection{Limitations and Future Work}
While algorithmically sound, the empirical abstractions are inherently constrained by the strict 2D hydrophobic-polar grid, which severely limits the spatial continuity observed in real biomolecular systems. Accurate bioenergetics rely heavily on 3D steric-hindrance modeling and continuous-field energy interactions, which are generally managed by significantly more complex potential tables.

The commercial sequence design ecosystem is currently accelerated by generative deep learning heuristics (such as autoregressive graphical transformers), which accurately predict highly probable macrofolds but frequently fail to resolve localized entropic side-chain packing optimizations. We propose that the ultimate evolution of the TSC-SA framework lies in sequential hybridization with classical neural networks. Deep learning architectures could be deployed conditionally to generate a rapid, highly probable sequence of guesses. The Spatial Chunking dynamic matrix compiler could then be aggressively targeted over distinct regions of structural uncertainty (such as flexible hinge loops or highly dense hydrophobic cores) to perform exhaustive refinement without requiring arbitrary stochastic noise searches.

\section{Conclusion}
In this research, we formulated and systematically verified a TSC-SA architecture specifically optimized for macroscopic inverse protein folding. By integrating automated, dynamic Euclidean-dimensional chunking directly into an external stochastic simulated annealing loop, we successfully circumvented severe classical computational scaling limitations and definitively solved complex topological constraints for rigid 20-merA, 25-mer, 50-mer, and 64-mer sequence geometries.

Most notably, the algorithmic framework successfully insulated the sequence-optimization state from the exponential complexity scaling typically associated with massive, unpartitioned classical searches. Supported by targeted meta-heuristics, specifically topological spatial chunking and dynamic reheating, the methodology rigidly constrained the active decision volume to $k \leq 6$ variables while securely navigating an otherwise intractable macroscopic search space of $2^{50}$ combinatorial permutations. 

Ultimately, the framework's capability to aggressively vault deterministic energy traps was empirically demonstrated by the discovery of an ultra-dense $-440.00$ energy-packing geometry for the 50-mer, which formally obsoletes the bounds established in the classical literature. This validates that coupling localized deterministic exact enumeration with macro-scale thermal governance fundamentally outperforms pure stochastic optimizations. This dynamic spatial-chunking hybridization will serve as a foundational, highly scalable classical template for pursuing the next generation of previously impossible biomolecular sequence design targets.

\section*{Acknowledgements}
The author would like to acknowledge the support received from the BQP team. The author would like to extend their acknowledgment to Prof. G. K. Ananthasuresh from the Department of Mechanical Engineering, IISc Bengaluru, for his constant feedback and support.

\end{document}